\documentclass[twocolumn,10pt]{IEEEtran}
\usepackage{braket}

\usepackage{hyperref}
\hypersetup{colorlinks,urlcolor=black}
\usepackage{booktabs}

\input{def.tex}
\usepackage{dsfont}
\usepackage{minibox}
\DeclareSymbolFont{matha}{OML}{txmi}{m}{it}
\DeclareMathSymbol{\varv}{\mathord}{matha}{118}
\usepackage{eurosym}
\usepackage{hhline,physics}

\usepackage{multicol}
\usepackage{algorithm}
\usepackage{graphicx}
\usepackage{textcomp}
\usepackage{pifont}
\usepackage{float}

\usepackage{subcaption}
\usepackage[multiple]{footmisc}

\usepackage{algorithmic}
\usepackage{tikz}
\usetikzlibrary{arrows.meta, positioning, calc}
\usetikzlibrary{calc,math}

\makeatletter
\renewcommand{\baselinestretch}{0.9}

\IEEEoverridecommandlockouts

\begin{document}
	\title{Fundamental Outage and Diversity Limits of Fluid Antenna Multiple Access Under Correlated Nakagami-$m$ Fading} 
\author{
	Anastasios~Papazafeiropoulos,~\IEEEmembership{Senior Member,~IEEE} 		

\thanks{Anastasios Papazafeiropoulos is with the Communications and Intelligent Systems Research Group, University of Hertfordshire, Hatfield AL10 9AB, U. K. (e-mail:tapapazaf@gmail.com). \par 
}
}
	\maketitle\vspace{-1.9cm}

\begin{abstract}
Existing analytical studies of 	fluid antenna multiple access (FAMA) are largely confined to Rayleigh fading, independent-port assumptions, or asymptotic selection metrics, leaving the fundamental outage behavior under realistic fading and spatial correlation conditions uncharacterized. 
	This paper fills this gap by presenting the first exact and asymptotically tight outage analysis of interference-limited FAMA systems under spatially correlated Nakagami-$m$ fading. By explicitly modeling the joint statistics of the desired and interference powers across fluid-antenna ports, \textcolor{black}{we first formulate the outage probability in terms of the joint
			desired-signal and interference distributions and then derive an explicit
			integral expression under a tractable single-reference-port spatial
			correlation model.} Building on this result, we establish a rigorous diversity-order characterization 
	revealing that the achievable diversity is jointly dictated by the fading severity and the spatial correlation rank imposed by the finite antenna aperture, rather than by the nominal number of ports. 	
	To enable accurate and tractable performance evaluation beyond the asymptotic regime, we further introduce a Gamma-matched outage approximation that preserves both the exact diversity order and the array-gain constant. The resulting closed-form expression provides an accurate characterization over a wide range of signal-to-interference ratio (SIR) thresholds and offers clear physical insight into the role of spatial correlation in FAMA systems. Extensive Monte Carlo (MC) simulations confirm the tightness of the exact analysis, the sharpness of the asymptotic results, and the accuracy of the proposed approximation. 
\end{abstract}

\section{Introduction}
Fluid antennas refer to a class of electronically reconfigurable radiating structures whose physical configuration or effective position can be altered in order to adapt their electromagnetic behavior, including polarization, resonant frequency, and radiation pattern, without relying on multiple permanently deployed antenna elements. Although the term fluid antenna was formally coined by Kar et al. in \cite{Kar2010}, the underlying concept of reconfigurable radiating media predates this work by several years. Early demonstrations include the use of liquid conductors to realize tunable antenna geometries, such as the mercury-based circular microstrip patch reported in \cite{Kosta2004}. Building on this idea, subsequent studies investigated a wide range of fluidic and fluid-inspired antenna implementations, including liquid-metal structures \cite{Hayes2012,Morishita2013,Dey2016} and antennas employing conductive or ionized solutions, such as sodium chloride or potassium chloride mixtures \cite{BordaFortuny2017,BordaFortuny2018,Singh2019}. The diversity potential of fluid antennas in single-user communication scenarios was first examined in \cite{FAMA_original,FluidAntenna1}, where outage probability performance was characterized, while the port-selection problem was further analyzed in \cite{FluidAntenna1}. In \cite{Papazafeiropoulos2026}, a novel joint SIM–FAS communication model in which transmission and reception are performed by a stacked intelligent metasurface and a fluid antenna system  has been recently proposed. Broader overviews of fluid and reconfigurable antenna technologies can be found in \cite{Paracha2019,Huang2021}, which collectively indicate that fluid-antenna concepts have progressed beyond proof-of-concept demonstrations and are increasingly recognized as viable components of future wireless systems \cite{Tariq2020}.

More recently, the fluid antenna paradigm has been extended to multiuser environments through the introduction of fluid antenna multiple access (FAMA) \cite{FAMA_MultipleAccess1}. In FAMA, each user dynamically repositions its antenna within a confined aperture and opportunistically selects the port that maximizes its instantaneous performance. This mechanism introduces a form of micro-scale spatial diversity that is fundamentally distinct from conventional multiple-input multiple-output (MIMO) techniques. Whereas classical MIMO systems exploit macro-scale spatial separation and rely on simultaneous observations across multiple radio-frequency (RF) chains, a fluid antenna operates using a single RF chain that sequentially probes multiple radiation points within a compact physical region. As a result, the diversity gains in FAMA stem from opportunistic port switching driven by small-scale channel variations, rather than from coherent multi-antenna combining.

Existing analytical treatments of FAMA are often limited in scope, typically assuming independent ports, focusing on single-input single-output (SISO) configurations, or characterizing only the statistics of the strongest port.
To address practical deployment aspects, Wong et al. proposed slow FAMA (s-FAMA) in \cite{Wong2023a}, where users update their selected antenna ports only upon significant channel variations. Subsequently, fast FAMA (f-FAMA) was introduced in \cite{Wong2023b}, enabling symbol-level port selection based solely on local received signal measurements, thereby supporting massive connectivity without the need for base-station precoding or multiuser detection.\footnote{\textcolor{black}{In this work, we focus on the slow-FAMA operating regime, where the receiver selects its antenna port based on the channel and interference conditions over a given coherence interval, and the selected port is kept fixed until a subsequent port update is performed. This setting is particularly relevant when port switching and channel probing cannot be performed at the symbol level due to hardware, signaling, or latency constraints. Therefore, the analytical framework developed in this paper characterizes the outage performance of slow FAMA under spatially correlated Nakagami-$m$ fading. The extension to fast FAMA, where the selected port may vary at the symbol level, would require an additional temporal switching and probing model and is left for future work.}} Furthermore, an orthogonal frequency-division multiplexing (OFDM)-based framework has recently been developed to integrate FAMA into 5G New Radio downlink systems, incorporating  signal-to-interference-plus-noise  ratio (SINR)-driven port selection, achievable-rate analysis, and link-level simulations that demonstrate substantial multiplexing gains over conventional fixed-antenna user equipment \cite{Hong2025}.  In \cite{Papazafeiropoulos2026a}, a unified analytical framework for
the multiple-input single-output (MISO) downlink with FAMA users served via maximum ratio transmission
(MRT) or zero-forcing (ZF) has been developed.

In parallel, Nakagami-$m$ fading has attracted considerable attention as a channel model despite its increased analytical complexity compared to Rayleigh or Rician fading, primarily due to its enhanced flexibility and generality in characterizing a wide range of wireless propagation conditions. In particular, the Nakagami-$m$ distribution provides a unified framework capable of modeling both weak and severe fading environments through a single shape parameter. Rayleigh fading emerges as a special case when $m=1$, while scenarios with a dominant line-of-sight component can be well approximated for $m>1$. Owing to this versatility, the Nakagami-$m$ model is especially suitable for describing the compounded fading effects encountered in reconfigurable intelligent surface (RIS)--assisted links, including double-fading phenomena~\cite{Selimis2021}. Nevertheless, analytical investigations of fluid antenna systems (FAS) under Nakagami-$m$ fading remain scarce, largely because the derivation of tractable closed-form expressions is significantly more challenging than under simpler fading assumptions. Tlebaldiyeva et al. developed an exact analytical framework for a single-antenna, $N$-port FAS operating over spatially correlated Nakagami-$m$ fading, deriving the joint probability density function (PDF)/cumulative distribution function (CDF) and an exact outage probability expression for selection combining across ports \cite{Tlebaldiyeva2022}.
Vega-S\'{a}nchez et al. built on this model by introducing an asymptotic-matching (Gamma) approximation that yields a simple closed-form outage expression and reveals a diversity order of $mN$ for correlated Nakagami-$m$ FAS, greatly reducing computational complexity \cite{VegaSanchez2024}. Moreover, the analysis in \cite{Huangfu2026} is confined to single-link Rician fading scenarios and focuses primarily on signal-strength maximization, without considering the joint effects of generalized fading severity, interference statistics, or correlation-induced diversity saturation. 
\textcolor{black}{It is also important to distinguish the present work from prior single-user Nakagami-$m$ FAS studies. Existing FAS analyses under Nakagami-$m$ fading mainly focus on noise-limited or single-link selection, where the selected port is the one with the strongest desired-signal power. Such models do not capture the defining feature of FAMA, namely interference avoidance through port-domain SIR selection. In an interference-limited FAMA network, the best port is not necessarily the port with the largest desired-signal gain, but the port that offers the most favorable desired-signal-to-interference ratio. Hence, the outage probability depends on the joint behavior of the desired and interfering links across the fluid-antenna ports. This motivates the exact and asymptotic SIR-outage analysis developed in this paper.}
\subsection*{Motivation and Scope}

To date, there has been no unified analytical treatment of fluid antenna multiple access (FAMA) operating under spatially correlated Nakagami-$m$ fading. Although Nakagami-$m$ fading is a flexible model that includes Rayleigh fading as the special case $m=1$ and can approximate milder line-of-sight-like propagation for $m>1$, its integration into FAMA analysis remains limited due to the challenges associated with correlated Gamma random processes.

Existing studies on Nakagami-$m$ FAS have mainly considered single-user or single-link settings, often assuming independent ports or focusing on strongest-port statistics. However, these results do not directly apply to FAMA, where port selection is based on the instantaneous SIR and therefore depends jointly on the desired-signal and aggregate-interference powers across correlated ports. Consequently, the outage behavior of interference-limited FAMA under spatial correlation and generalized fading severity remains insufficiently characterized.

Motivated by this gap, this paper develops an analytical framework for the outage performance of FAMA receivers over spatially correlated Nakagami-$m$ fading channels. We derive an exact outage probability expression that accounts for the joint statistics of the desired and interfering signals across the fluid-antenna ports. We then establish the high-reliability asymptotic behavior and show that the effective diversity order is given by
\[
d_{\mathrm{eff}}=m_s\operatorname{rank}(\mathbf R),
\]
where $\mathbf R$ is the spatial correlation matrix induced by the finite aperture. To enable tractable evaluation beyond the asymptotic regime, we further introduce a Gamma-matched approximation that preserves both the diversity order and the array-gain behavior, yielding a simple closed-form expression accurate over a wide range of SIR thresholds.

\textcolor{black}{The use of Nakagami-$m$ fading is not merely a mathematical extension of Rayleigh-based FAMA analysis. It enables the desired-link fading parameter $m_s$ and the interference-link fading parameter $m_I$ to be varied independently, revealing their distinct engineering roles. Specifically, $m_s$ determines the asymptotic diversity exponent through $d_{\mathrm{eff}}=m_s\operatorname{rank}(\mathbf R)$, whereas $m_I$ mainly affects the finite-SIR outage level and the associated array-gain constant. Hence, the proposed analysis identifies not only whether FAMA provides selection diversity, but also when this diversity is limited by aperture-induced correlation and when the outage behavior is dominated by interference fluctuations.}

\textcolor{black}{The novelty of this work is also distinct from classical correlation-limited diversity results in conventional MIMO systems. Unlike conventional receive diversity, where multiple antenna branches are observed simultaneously and combined coherently or noncoherently, a FAMA receiver uses a single RF chain and selects one port according to an instantaneous SIR criterion. Thus, the relevant outage event depends on the joint condition,	which couples the correlated desired-signal and aggregate-interference powers across all ports. Therefore, while it is well known that spatial correlation limits diversity, this paper establishes how this principle appears in interference-limited FAMA under correlated Nakagami-$m$ fading and quantifies the separate roles of fading severity, interference statistics, and aperture-induced spatial correlation.}

\subsection*{Main Contributions}

The main contributions of this paper can be summarized as follows:

\begin{itemize}
	
%
	\item \textcolor{black}{ \textbf{Exact SIR-outage analysis for interference-limited FAMA:}
		Unlike classical diversity analysis and single-user FAS studies, which mainly characterize the desired-signal gain or the strongest desired-signal port, this paper analyzes the SIR-based outage event of a FAMA receiver. The derived expression explicitly accounts for the joint statistics of the correlated desired-signal powers and the correlated aggregate-interference powers across the fluid-antenna ports. This provides an exact outage characterization for slow-FAMA under spatially correlated Nakagami-$m$ fading without relying on independent-port assumptions.}
	
\textcolor{black}{\item \textbf{Diversity-order characterization for SIR-based FAMA selection:}
	We show that the effective diversity order of the interference-limited FAMA receiver is
	\[
	d_{\mathrm{eff}}=m_s\operatorname{rank}(\mathbf R).
	\]
	Although correlation-limited diversity is a known principle in conventional multi-antenna systems, the present result establishes its precise form for SIR-based FAMA port selection under Nakagami-$m$ fading. It also separates the role of the desired-link fading severity, which determines the outage slope, from the role of the interference statistics, which affect the array-gain constant and the finite-SIR outage level.}
	
	\item \textbf{Gamma-matched outage approximation with asymptotic accuracy:}
	Motivated by the derived power-law behavior, we propose a Gamma-matched approximation for the outage probability of the maximum SIR. The approximation preserves both the exact diversity order and the array-gain constant, yielding a compact closed-form expression in terms of the incomplete Gamma function. Numerical results demonstrate that this approximation remains highly accurate not only in the asymptotic high-SIR regime but also over a broad range of practical SIR thresholds.
	
	\item \textcolor{black}{\textbf{Engineering design implications for FAMA deployment:}
	The analytical and numerical results are translated into concrete design guidelines for fluid-antenna receivers. First, increasing the number of selectable ports improves reliability only when the aperture and port spacing create additional effective spatial degrees of freedom. Otherwise, the diversity gain saturates because it is governed by $\operatorname{rank}(\mathbf R)$ rather than by the nominal number of ports. Second, the Nakagami parameter of the desired link, $m_s$, controls the asymptotic outage slope, while the interference fading parameter, $m_I$, strongly influences the finite-SIR outage level. This distinction shows that practical FAMA design should not only seek strong desired-signal locations, but should also exploit port positions where the interference exhibits sufficiently large fluctuations and deep fades. These observations provide direct guidance on aperture sizing, port-density selection, and the operating scenarios in which FAMA is expected to provide substantial reliability gains. Moreover, the proposed correlated FAMA scheme is  compared with independent-port FAMA, random-port selection, and desired-signal-based selection, thereby clarifying the individual effects of spatial correlation and interference-aware port selection.}
	
\end{itemize}

The remainder of the paper is structured as follows. 
Section~\ref{system} presents the considered interference-limited FAMA system model, detailing the fluid-antenna architecture, the spatially correlated Nakagami-$m$ channel model, and the definition of the per-port and maximum SIR under FAMA port selection. 
Section~\ref{statistical} develops an exact analytical characterization of the outage probability by deriving the joint statistics of the desired and interfering signals across correlated ports, resulting in a closed-form integral expression for the FAMA outage probability. 
In Section~\ref{fama}, the high-reliability regime is investigated, where asymptotic outage expressions are obtained and the effective diversity order is rigorously characterized as a function of the Nakagami fading parameter and the rank of the spatial correlation matrix. This section also introduces a Gamma-matched approximation that preserves the asymptotic behavior while enabling a compact and tractable outage representation. 
Section~\ref{Numerical} provides extensive numerical results that validate the exact analysis, the asymptotic diversity characterization, and the Gamma-matched approximation, while offering insights into the impact of fading severity, spatial correlation, and system dimensionality on the outage performance of FAMA. 
Finally, Section~\ref{conclusion} concludes the paper.

\paragraph*{Notation}
Boldface uppercase letters denote matrices, while boldface lowercase letters
represent vectors.
For a matrix $\mathbf{X}$, the operators $\mathbf{X}^{\mathsf{T}}$ and
$\mathbf{X}^{\mathsf{H}}$ indicate the transpose and Hermitian transpose,
respectively.
The expectation and variance operators are denoted by $\mathbb{E}\{\cdot\}$
and $\mathrm{Var}\{\cdot\}$.
The function $J_0(\cdot)$ represents the zeroth-order Bessel function of the
first kind, and $\mathbf{I}_M$ denotes the $M \times M$ identity matrix.

\section{System Model}
\label{system}

\textcolor{black}{We consider an interference-limited downlink network in which a serving base station communicates with a typical user equipment (UE), while multiple cochannel transmitters or users share the same time--frequency resource and generate interference. The typical UE is equipped with a single-RF-chain FAS that operates according to the slow-FAMA principle. Specifically, the UE probes the available fluid-antenna ports over a given coherence interval, selects the port that provides the most favorable SIR, and keeps this port fixed until the next port-selection update. }
	\begin{figure}%
	\centering
	\includegraphics[width=\linewidth]{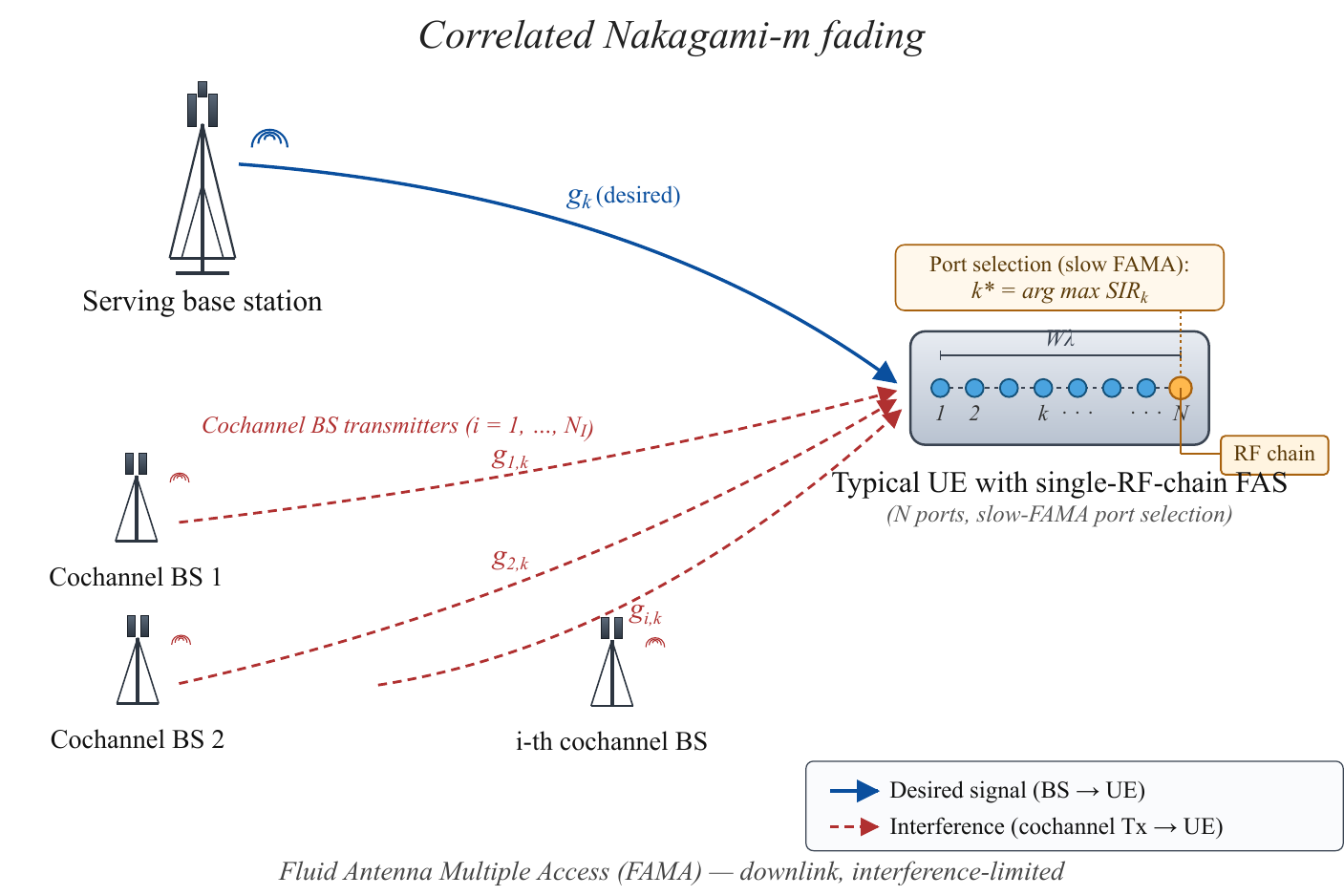}
	\caption{\textcolor{black}{A slow-FAMA system model under spatially correlated Nakagami-$m$ fading.}}
	\label{fig0}
\end{figure}
\subsection{Fluid Antenna Architecture}

Each user terminal employs a fluid antenna with $N$ selectable ports uniformly
distributed along a one-dimensional aperture of normalized length $W\lambda$,
where $\lambda$ denotes the carrier wavelength. The $k$-th port is located at
\begin{equation}
	d_k = \frac{k-1}{N-1} W\lambda, \qquad k = 1,\ldots,N.
\end{equation}
Following common
practice, each port is modeled as an ideal point antenna to facilitate
analytical tractability.

\subsection{Signal and Interference Model}

We focus on a typical user and omit the user index for clarity. The complex
baseband signal received at the $k$-th port is given by
\begin{equation}
	y_k = g_k s + \sum_{i=1}^{N_I} g_{i,k} s_i + n_k,
\end{equation}
where $s$ denotes the desired information symbol, $s_i$ is the symbol transmitted
by the $i$-th interferer, and $n_k \sim \mathcal{CN}(0,\sigma_n^2)$ is additive
white Gaussian noise (AWGN). The desired channel coefficient $g_k$ models small-scale fading between the
serving transmitter and the $k$-th port, while $g_{i,k}$ represents the channel
from the $i$-th interferer to the same port. In interference-dominated regimes,
the aggregate interference at port $k$ can be expressed as
\begin{equation}
	g_{I,k} \triangleq \sum_{i=1}^{N_I} g_{i,k} s_i,
\end{equation}
which is treated as a single equivalent fading coefficient \cite{Simon2004}.

\subsection{Nakagami-$m$ Fading With Spatial Correlation}

\subsubsection{Desired-Signal Fading Model}

The small-scale fading amplitude of the desired link at the $k$-th fluid-antenna
port is denoted by $|g_k|$ and is modeled as a Nakagami-$m_s$ random variable.
Its PDF is given by
\begin{equation}
	f_{|g_k|}(r)
	=
	\frac{2 m_s^{m_s}}{\Gamma(m_s)\Omega_s^{m_s}}
	r^{2m_s-1}
	\exp\!\left(-\frac{m_s}{\Omega_s} r^2\right),
	\qquad r \ge 0,
	\label{eq:nakagami_pdf}
\end{equation}
where $m_s \ge \tfrac{1}{2}$ is the fading severity parameter and
$\Omega_s = \mathbb{E}[|g_k|^2]$ denotes the average received power, which is
assumed identical across all ports.

Since the square of a Nakagami-$m$ envelope follows a Gamma distribution, the
corresponding channel power gain $S_k \triangleq |g_k|^2$ 
is Gamma distributed as
\begin{equation}
	S_k \sim \mathrm{Gamma}\!\left(m_s,\frac{\Omega_s}{m_s}\right),
	\label{eq:nakagami_gamma}
\end{equation}
where the first and second arguments denote the shape and scale parameters,
respectively.

\subsubsection{Interference-Signal Fading Model}
In general, different interfering transmitters may experience distinct
propagation conditions due to unequal distances, blockage, or scattering
environments. To capture this effect, we allow each interferer--to--port link
to follow an individual Nakagami-$m$ distribution, i.e., the corresponding power is modeled as
\begin{equation}
	|g_{i,k}|^2 \sim \mathrm{Gamma}\!\left(m_{I,i},\frac{\Omega_{I,i}}{m_{I,i}}\right),
\end{equation}
where $m_{I,i}$ and $\Omega_{I,i}$ denote the fading parameter and the average
received interference power of the $i$-th interferer, respectively.

\textcolor{black}{Due to the compact aperture of the fluid antenna, the channel
gains observed across ports are spatially correlated. In general,
the spatial dependence among the $N$ ports is described by a
correlation matrix $\mathbf R\in\mathbb C^{N\times N}$, whose
entries depend on the aperture geometry, port locations, antenna
radiation pattern, and angular distribution of the incoming waves.
The analytical results in this paper are expressed in terms of
$\mathbf R$ and its effective rank, which captures the number of
spatial degrees of freedom supported by the finite aperture.}

\textcolor{black}{For analytical tractability and to obtain explicit expressions, we adopt
	the classical one-dimensional isotropic-scattering model together with a
	single-reference-port representation. Specifically, port~1, located at
	the first endpoint of the fluid-antenna aperture, is selected as the
	reference port. With uniformly spaced ports, the distance between port~1
	and port~$k$ is $(k-1)W/(N-1)$. Hence, the correlation coefficient
	between the reference port and the $k$-th port is
	\begin{equation}
		\mu_k
		=
		J_0\!\left(
		2\pi W\frac{k-1}{N-1}
		\right),
		\qquad k=1,\ldots,N,
		\label{eq:mu_k_uniform}
	\end{equation}
	where $\mu_1=1$ by definition.}

\textcolor{black}{Equivalently, the corresponding full correlation matrix can be
	written as
	\begin{equation}
		[\mathbf R]_{i,j}
		=
		J_0\left(2\pi W\frac{|i-j|}{N-1}\right),
		\quad i,j=1,\ldots,N.
	\end{equation}
	This Bessel/Jakes-type model is widely used as a tractable
	baseline for one-dimensional apertures under isotropic
	scattering. Nevertheless, we emphasize that simplified
	single-reference or constant-correlation representations may
	overestimate the gain of FAS/FAMA when the ports are densely
	sampled within a finite aperture. Recent work has shown that
	more accurate spatial correlation models, such as the
	block-correlation model in~\cite{Ramirez2024BlockCorrelation},
	can better capture the saturation of spatial degrees of freedom
	and the resulting performance loss in dense FAS deployments \cite{Hong2026FAMAReview,Hong2026FASSurvey,New2025FASTutorial}.
	Therefore, in the numerical section, we also include a
	block-correlated FAMA benchmark to quantify the impact of
	realistic spatial correlation.}

The aggregate interference power at the $k$-th port is given by
\begin{equation}
	I_k = \sum_{i=1}^{N_I} |g_{i,k}|^2.
\end{equation}
When the interferers have non-identical parameters
$\{(m_{I,i},\Omega_{I,i})\}$, $I_k$ is the sum of independent but
non-identically distributed Gamma random variables and therefore does not
admit an exact Gamma distribution. For analytical tractability, we approximate
$I_k$ by a Gamma random variable using moment matching \cite{Simon2004}. Specifically, the first two moments of $I_k$ are
\begin{align}
	\mathbb{E}[I_k] &= \sum_{i=1}^{N_I} \Omega_{I,i}, \quad
	\mathrm{Var}[I_k] = \sum_{i=1}^{N_I} \frac{\Omega_{I,i}^2}{m_{I,i}}.
\end{align}
Matching these moments to a Gamma distribution yields the effective parameters
\begin{equation}
	m_{I,\mathrm{eq}}
	=
	\frac{\left(\sum_{i=1}^{N_I} \Omega_{I,i}\right)^2}
	{\sum_{i=1}^{N_I} \Omega_{I,i}^2 / m_{I,i}},
	\qquad
	\Omega_{I,\mathrm{eq}}
	=
	\sum_{i=1}^{N_I} \Omega_{I,i},
\end{equation}
such that
\begin{equation}
	I_k \approx \mathrm{Gamma}\!\left(m_{I,\mathrm{eq}},
	\frac{\Omega_{I,\mathrm{eq}}}{m_{I,\mathrm{eq}}}\right).
\end{equation}
This approximation becomes exact when all interferers are statistically
identical and is widely used to characterize sums of heterogeneous Gamma
random variables \cite{Simon2004}.

\textcolor{black}{\subsection{Spatial Block-Correlation Benchmark}}

\textcolor{black}{	To account for more realistic FAS channel behavior, we also
	consider the spatial block-correlation model of
	\cite{Ramirez2024BlockCorrelation} as a benchmark. The key
	idea is to approximate a realistic spatial correlation matrix,
	such as the one generated by Jakes' or Clarke's model, by a
	block-diagonal matrix. Ports within the same spatial coherence
	block are strongly correlated, whereas ports belonging to
	different blocks are treated as approximately independent. This
	model captures the fact that, when many ports are placed within
	a finite aperture, nearby ports do not create independent fading
	observations.}

\textcolor{black}{	Let the $N$ ports be divided into $B$ spatial blocks
	$\{\mathcal B_1,\ldots,\mathcal B_B\}$, where block
	$\mathcal B_b$ contains $L_b$ ports and
	$\sum_{b=1}^{B}L_b=N$. The block-correlation matrix is written
	as
	\begin{equation}
		\mathbf R_{\mathrm{blk}}
		=
		\operatorname{blkdiag}
		\left(
		\mathbf R_1,\ldots,\mathbf R_B
		\right),
	\end{equation}
	where each block is modeled as
	\begin{equation}
		\mathbf R_b
		=
		(1-\rho_b)\mathbf I_{L_b}
		+
		\rho_b\mathbf 1_{L_b}\mathbf 1_{L_b}^{T},
		\quad b=1,\ldots,B.
	\end{equation}
	Here, $\rho_b\in[0,1)$ denotes the intra-block correlation
	coefficient, while different blocks are independent. The number
	of blocks and the intra-block correlations can be selected to
	match the dominant eigenvalue structure of a target correlation
	matrix. In this sense, $\mathbf R_{\mathrm{blk}}$ provides a
	tractable approximation to fully spatially correlated FAS
	channels while preserving the main effect of finite-aperture
	correlation, namely the saturation of the number of effective
	spatial degrees of freedom.}

\subsection{SIR and FAMA Port Selection}

In interference-limited scenarios, thermal noise is negligible compared to the
aggregate interference power. Accordingly, the instantaneous SIR at the $k$-th port is
\begin{equation}
	\mathrm{SIR}_k \triangleq \frac{|g_k|^2}{|g_{I,k}|^2}.
\end{equation}

The fluid antenna selects the port that maximizes the SIR, i.e., we have 
\begin{equation}
	\mathrm{SIR}_{\max} = \max_{k=1,\ldots,N} \mathrm{SIR}_k,\label{eq13}
\end{equation}
which corresponds to exploiting spatial locations where the interference
experiences deep fading while the desired signal remains sufficiently strong.\footnote{
\textcolor{black}{This selection rule corresponds to slow FAMA, where the port maximizing the instantaneous SIR is selected for the considered channel realization or coherence interval. Unlike fast FAMA, which may update the selected port at the symbol level, the present model focuses on outage performance under quasi-static port selection. This distinction is important because the derived outage probability characterizes the reliability achieved after port selection within a coherence interval, rather than the symbol-by-symbol switching behavior of fast FAMA.}}

\textcolor{black}{	\subsection{Outage-Constrained Multiplexing Interpretation}}

\textcolor{black}{Although the main focus of this paper is the outage performance
	of a typical FAMA receiver, the derived outage probability also
	provides a direct way to quantify the multiplexing capability of
	the network. In FAMA, multiple users may share the same
	time--frequency resource, while each receiver exploits port
	selection to find a spatial position where the desired signal is
	sufficiently strong relative to the aggregate interference. Hence,
	the number of simultaneously active cochannel users that can be
	supported is directly related to the outage probability under a
	given SIR requirement.}

\textcolor{black}{	Let $\epsilon$ denote a target outage probability and let
	$N_I$ denote the number of active interfering users. For a given
	SIR threshold $\gamma$, we define the outage-constrained
	multiplexing capability as
	\begin{equation}
		N_I^{\star}(\gamma,\epsilon)
		=
		\max
		\left\{
		N_I:
		P_{\mathrm{out}}(\gamma;N_I)\leq \epsilon
		\right\}.
	\end{equation}
	This metric represents the maximum number of cochannel
	interferers that can be tolerated while maintaining the target
	outage constraint. Equivalently, the number of simultaneously
	active users supported on the same time--frequency resource is
	$N_I^{\star}+1$, including the desired transmitter.}
	
	\textcolor{black}{The proposed outage analysis therefore provides a reliability-
	based characterization of the multiplexing gain of FAMA. A
	larger number of effective independent ports allows the receiver
	to better exploit spatial interference fluctuations, thereby
	increasing $N_I^{\star}$. Conversely, strong spatial correlation
	reduces the effective rank of the port correlation matrix and
	limits the number of cochannel users that can be supported
	under the same outage constraint.}

\section{Outage Probability Analysis}
\label{statistical}

For a target SIR threshold $\gamma$, an outage event occurs when
\begin{equation}
	\mathcal{O}_{\mathrm{SIR}}
	=
	\left\{
	\mathrm{SIR}_{\max} < \gamma
	\right\}
	=
	\left\{
	|g_k|^2 < \gamma |g_{I,k}|^2,\ \forall k
	\right\}.
\end{equation}
The main objective of this paper is to characterize the outage probability
$\Pr(\mathcal{O}_{\mathrm{SIR}})$ under correlated Nakagami-$m$ fading and to
analyze how it scales with the number of ports, the fading parameters, and the
spatial correlation induced by the fluid antenna aperture.

Based on the above outage definition, the following theorem provides an exact
characterization of the outage probability for the FAMA receiver under
correlated Nakagami-$m$ fading. The exact outage probability is derived by conditioning on the aggregate
interference power and exploiting the conditional distribution of the desired
signal under correlated Nakagami-$m$ fading. This leads to an integral
representation involving the joint statistics of the desired and interfering
channels, as stated in Theorem~1. The resulting expression captures the impact
of spatial correlation across fluid-antenna ports through the correlation
coefficients induced by the finite aperture.
\begin{theorem}
	\label{Theorem1}
The outage probability of the FAMA receiver with respect to a target SIR
	threshold $\gamma>0$ is given by
	\begin{align}
	P_{\mathrm{out}}(\gamma)
	&=\!\!\!
	\int_{0}^{\infty}\!\!
	f_{Y_1}(z)
	\int_{0}^{\gamma z}
	f_{X_1}(t)
	\prod_{k=2}^{N}
	\Psi_k(t,z;\gamma)
	\, dt\, dz,\label{pout}
\end{align}
where 
\begin{align}
		f_{X_1}(t)
&	=
	\frac{m_s^{m_s}}{\Gamma(m_s)}
	t^{m_s-1} e^{-m_s t},\\
		f_{Y_1}(z)
	&=
	\frac{m_I^{m_I}}{\Gamma(m_I)}
	z^{m_I-1} e^{-m_I z},
\end{align}
	\begin{equation}
		\Psi_k(t,z;\gamma)
		=
		\underbrace{\Psi_k^{(1)}(t,z;\gamma)}_{\text{Rayleigh baseline}}
		-
		\underbrace{\Psi_k^{(2)}(t,z;\gamma)}_{\text{Nakagami correction}} .
	\end{equation}
	with \begin{align}
		&\Psi_k^{(1)}(t,z;\gamma)
		=
		\int_{0}^{\infty}
		\Bigg[
		1-
		Q_{1}\!\left(
		\sqrt{\alpha_k t},\,\sqrt{\beta_k\gamma\,y}
		\right)
		\Bigg]\,
		\frac{m_I}{1-\mu_k^2}
				\nn\\
		&\times\left(\frac{y}{\mu_k^2 z}\right)^{\frac{m_I-1}{2}}
		\exp\!\left(
		-\frac{m_I}{1-\mu_k^2}(y+\mu_k^2 z)
		\right)	\nn\\
		&\times
		I_{m_I-1}\!\left(
		\frac{2m_I\mu_k}{1-\mu_k^2}\sqrt{yz}
		\right)\,dy,
				\label{eq:Psi1_explicit}
	\end{align}
	and
		\begin{align}
		&	\Psi_k^{(2)}(t,z;\gamma)
		=
		\left(
		\frac{m_I\gamma}{m_s(1-\mu_k^2)}
		\right)^{m_s}
		\exp\!\left(
		-\frac{m_I\gamma z + m_s t}{1-\mu_k^2}
		\right)
		\nn\\
		&\times
		\sum_{\ell=1}^{m_s-1}
		\frac{1}{\ell!}
		\left(
		\frac{m_I\gamma\mu_k^2}{1-\mu_k^2}z
		\right)^{\ell}
		I_{m_I+\ell-1}
		\!\left(
		\frac{2\mu_k\sqrt{m_I m_s\gamma z t}}{1-\mu_k^2}
		\right),		\label{eq:psi2_closed_form}
	\end{align}
\end{theorem}
\begin{proof}
	Please see Appendix~\ref{Theorem1proof}.	
\end{proof}
Note that the term ``Rayleigh baseline'' refers to the first-order Marcum-$Q$ structure
associated with Rayleigh desired-signal fading ($m_s=1$), while the
interference statistics remain Nakagami-$m_I$ distributed.

\begin{remark}[Consistency with Rayleigh Fading]
	\label{rem:rayleigh_special_case}
	The general expression in Theorem~\ref{Theorem1} reduces to the classical
	Rayleigh fading case by setting $m_s=m_I=1$ \cite[Eq. 17]{FAMA_MultipleAccess1}. In this case, the desired-signal
	and interference powers are exponentially distributed, i.e.,
	$f_{X_1}(t)=e^{-t}$ and $f_{Y_1}(z)=e^{-z}$. Moreover, the Nakagami correction
	term $\Psi_k^{(2)}(t,z;\gamma)$ vanishes since the finite sum in
	\eqref{eq:psi2_closed_form} becomes empty. As a result,
	$\Psi_k(t,z;\gamma)=\Psi_k^{(1)}(t,z;\gamma)$, and the outage probability
	expression collapses to the Rayleigh baseline characterized solely by the
	first-order Marcum-$Q$ function $Q_1(\cdot,\cdot)$.
\end{remark}

\textcolor{black}{\emph{Extension to two-dimensional FAS:}
	Although the system model adopts a one-dimensional FAS for analytical clarity, the main diversity and outage conclusions extend naturally to two-dimensional FAS deployments. In a 2D-FAS, the $N$ ports are distributed over a planar aperture rather than along a line. \textcolor{black}{Let the position of the $k$-th port be denoted by
		\begin{equation}
			\mathbf p_k=[p_{x,k},p_{y,k}]^T.
		\end{equation}
		Under isotropic scattering, the spatial correlation between ports $i$
		and $j$ may be expressed as a function of their Euclidean separation,
		for example
		\begin{equation}
			[\mathbf R_{\mathrm{2D}}]_{i,j}
			=
			J_0\left(
			2\pi\frac{\|\mathbf p_i-\mathbf p_j\|}{\lambda}
			\right),
			\quad i,j=1,\ldots,N.
	\end{equation}}
	More generally, $\mathbf R_{\mathrm{2D}}$ can be generated from any physically motivated 2D spatial correlation model that accounts for the aperture geometry, angular scattering distribution, and radiation pattern.}

\textcolor{black}{Once the correlation matrix is specified, the asymptotic diversity result remains governed by the effective number of spatial degrees of freedom supported by the aperture. In particular, for a 2D-FAS, the diversity order becomes
	\begin{equation}
		d_{\mathrm{eff}}^{\mathrm{2D}}
		=
		m_s\operatorname{rank}(\mathbf R_{\mathrm{2D}}).
	\end{equation}
	Therefore, the nominal number of ports does not by itself determine the outage slope. A 2D-FAS can provide a larger effective rank than a 1D-FAS because the ports span an aperture area rather than a line segment, which may increase spatial decorrelation and improve outage performance. However, if the ports are densely placed within a small physical aperture, the rank or effective rank of $\mathbf R_{\mathrm{2D}}$ may still saturate, leading to diminishing returns. Hence, the conclusions derived for the 1D-FAS case scale to 2D-FAS by replacing the 1D correlation matrix with the corresponding 2D correlation matrix and interpreting the diversity gain in terms of its effective rank.}

\begin{remark}[Generality and scope of the correlation model]
		\label{rem:correlation_scope}
	\textcolor{black}{	For arbitrary desired-signal and interference correlation matrices
		$\mathbf R_s$ and $\mathbf R_I$, respectively, the outage probability
		admits the general representation
		\begin{equation}
			P_{\mathrm{out}}(\gamma)
			=
			\int_{\mathbb{R}_{+}^{N}}
			F_{\mathbf X}
			\left(
			\gamma y_1,\ldots,\gamma y_N;\mathbf R_s
			\right)
			f_{\mathbf Y}
			\left(
			y_1,\ldots,y_N;\mathbf R_I
			\right)
			\,\mathrm d\mathbf y,
			\label{eq:general_outage}
		\end{equation}
		where $F_{\mathbf X}(\cdot;\mathbf R_s)$ and
		$f_{\mathbf Y}(\cdot;\mathbf R_I)$ denote the joint CDF of the
		desired-signal powers and the joint PDF of the interference powers,
		respectively. Although \eqref{eq:general_outage} is valid for any
		physically admissible correlation structure for which these joint
		distributions exist, its explicit evaluation generally requires the
		complete multivariate distributions rather than only their pairwise
		correlation coefficients.}
		
		\textcolor{black}{In particular, the product representation in
		Theorem~\ref{Theorem1} relies on the conditional-independence structure
		of the adopted single-reference-port correlated Nakagami-$m$ model.
		Accordingly, it may be used with correlation coefficients generated by
		a different spatial model only when they remain consistent with this
		conditional construction. It cannot, in general, be extended to an
		arbitrary full correlation matrix $\mathbf R$ merely by replacing
		$\{\mu_k\}$ with its first-row entries. The Bessel/Jakes model adopted
		in this paper provides a physically motivated and analytically
		tractable specification of these reference-port coefficients.}
		
	\textcolor{black}{	The diversity-order result in Theorem~\ref{Theorem2} applies to
		correlation matrices admitted by the correlated Nakagami-$m$
		construction considered in this work, for which the effective number
		of spatial modes is characterized by $\operatorname{rank}(\mathbf R)$.
		It should not be interpreted as applying to an otherwise arbitrary
		multivariate fading distribution specified only through its
		correlation matrix.}
		
	\textcolor{black}{	The explicit product-form outage expression and the closed-form
		array-gain coefficient are therefore specific to the adopted
		reference-port construction. For another correlation structure, the
		corresponding expressions must be derived from the complete joint
		distributions in \eqref{eq:general_outage}. Accordingly, the
		block-correlation model introduced in Section~II-D and evaluated in
		Fig.~\ref{fig11} is used only as a numerical robustness benchmark.
		Its outage probability is obtained through Monte Carlo simulation by
		generating the desired-signal and interference channel realizations
		according to $\mathbf R_{\mathrm{blk}}$, computing the instantaneous
		SIR at every port, and applying the FAMA port-selection rule. }
\end{remark}

While the exact formulation provides a complete statistical description, it
does not immediately reveal how outage performance scales with key system
parameters. To gain deeper insight, we next study the asymptotic behavior of the
outage probability in the high-reliability regime.

\section{Asymptotic Outage Probability and Diversity Order}
\label{fama}

In this subsection, we characterize the high-reliability behavior of the
FAMA receiver by deriving the asymptotic outage probability and the
associated diversity order in the interference-limited regime.

Let $\mathbf{R}$ denote the spatial correlation matrix of the fluid-antenna
ports, induced by the finite aperture $W$ and characterized by the correlation
coefficients $\{\mu_k\}$ in \eqref{eq:mu_k_uniform}. Accordingly, the spatial correlation matrix $\mathbf{R} \in \mathbb{C}^{N\times N}$
is defined by
\begin{equation}
	R_{ij}
	=
	J_0\!\left(2\pi W \frac{|i-j|}{N-1}\right),
	\qquad i,j = 1,\ldots,N,
	\label{eq:R_toeplitz}
\end{equation}
such that $\mu_k = R_{1k}$ corresponds to the correlation coefficient between
the $k$-th port and the reference port.

 We next characterize the high-reliability behavior of the FAMA receiver by
studying the asymptotic decay of the outage probability as the target SIR
threshold $\gamma$ approaches zero.

\begin{theorem}[Asymptotic outage behavior and diversity order]
	\label{Theorem2}
Under spatially correlated Nakagami-$m$ fading and in the high-SIR regime $\gamma\to 0$, the outage probability
	admits the asymptotic representation
	\begin{equation}
		P_{\mathrm{out}}(\gamma)
		\;\sim\;
		c\,\gamma^{d_{\mathrm{eff}}},
		\qquad \gamma \to 0,\label{eq:asymp_outage1}
	\end{equation}
	where $c>0$ is a constant independent of $\gamma$.
Also, the effective diversity order of the FAMA receiver is given by
	\begin{equation}
		d_{\mathrm{eff}}
		=
		m_s~r\,   \label{eq:diversity_order}
	\end{equation}
	with $r=\mathrm{rank}(\mathbf{R})$.
\end{theorem}
\begin{proof}
	Please see Appendix~\ref{Theorem2proof}.	
\end{proof}

\subsubsection*{Discussion and Physical Interpretation}

The result in \eqref{eq:diversity_order} yields the following insights.

\begin{itemize}
	\item \emph{Correlation-limited diversity:}
	Although the fluid antenna employs $N$ selectable ports, the achievable
	diversity order depends on $\mathrm{rank}(\mathbf{R})$ rather than on
	$N$. Strong spatial correlation induced by a compact aperture reduces
	the number of effective diversity branches, leading to diminishing
	returns as $N$ increases.
	
	\item \emph{Interference robustness:}
In the interference-limited SIR regime, the asymptotic diversity exponent is
governed by the desired-signal fading severity and the spatial correlation
rank induced by the finite fluid-antenna aperture, yielding
$d_{\mathrm{eff}} = m_s r$.
In contrast, the statistics of the interfering signals affect only the
multiplicative array-gain constant and do not alter the diversity order.

	\item \emph{Consistency with classical selection combining:}
	For i.i.d.\ Rayleigh fading ($m_s=1$ and $\mathbf{R}=\mathbf{I}$),
	\eqref{eq:diversity_order} reduces to $d_{\mathrm{eff}}=N$, recovering
	the well-known diversity order of selection combining.
\end{itemize}

\begin{remark}
	The asymptotic result in \eqref{eq:asymp_outage1} complements the exact
	outage expression in Theorem~1 by explicitly revealing how spatial
	correlation and fluid-antenna geometry constrain the reliability gains
	of FAMA systems in the high-reliability regime.
\end{remark}

While Theorem~\ref{Theorem2} characterizes the asymptotic slope of the outage probability, a
more accurate and analytically tractable approximation can be obtained by
matching its asymptotic behavior to a parametric distribution.

Motivated by the power-law decay identified in Theorem~\ref{Theorem2}, we employ an
asymptotic matching approach based on the Gamma distribution. This yields a
closed-form approximation that preserves both the diversity order and the
array-gain characteristics of the FAMA receiver.

\begin{theorem}
	\label{Theorem3}
 The outage probability of the FAMA receiver with respect to a target SIR
	threshold $\gamma>0$  admits the following approximation
	\begin{equation}
		P_{\mathrm{out}}(\gamma)
		\;\approx\;
		\frac{\Upsilon\!\left(\alpha,\frac{\gamma}{\beta}\right)}{\Gamma(\alpha)},
		\qquad
		\alpha = m_s r,
		\qquad
		Li2012,
		\label{eq:gamma_match}
	\end{equation}
	where $\Upsilon(\cdot,\cdot)$ denotes the lower incomplete Gamma function.
	Moreover, the outage probability exhibits the asymptotic behavior
	\begin{equation}
		P_{\mathrm{out}}(\gamma)
		\sim
		a_0\,\gamma^{m_s r},
		\qquad
		\gamma\to 0,
		\label{eq:asymp_outage}
	\end{equation}
	with diversity order $m_s r$ and array-gain constant $a_0$ given in
	closed form by
	\begin{align}
		a_0
		&=
		\frac{m_I^{m_I}}{m_s\,\Gamma(m_I)}
		\left(
		\prod_{k\in\mathcal{K}_r}
		\frac{\beta_k^{m_s}}{2^{m_s}m_s!}
		\right)
		\nonumber\\
		&\quad\times
		\sum_{n_2=0}^{m_I-1}\cdots\sum_{n_N=0}^{m_I-1}
		\left(
		\prod_{k\in\mathcal{K}_r} A_{k,n_k}
		\right)\nonumber\\
			&\quad\times\frac{
			\Gamma\!\left(
			m_s+m_I+\sum_{k\in\mathcal{K}_r}n_k
			\right)
		}{
			\left(
			m_I+\sum_{k=2}^{N}c_k\mu_k^2
			\right)^{m_s+m_I+\sum n_k}
		},
		\label{eq:a0_final}
	\end{align}
	where $c_k=\frac{m_I}{1-\mu_k^2},$
		and 
		\begin{align}
				A_{k,n}
			=
			\frac{
				\Gamma\!\left(m_s+\frac{3m_I-1}{2}+n\right)
			}{
				n!\,\Gamma(m_I+n)
			}
			c_k^{-\left(m_s+\frac{m_I-1}{2}-n\right)}
			\mu_k^{2n}.
		\end{align}
\end{theorem}
\begin{proof}
	Please see Appendix~\ref{Theorem3proof}.	
\end{proof}
\textcolor{black}{The closed-form coefficient $a_0$ in
\eqref{eq:a0_final}, and hence the scale parameter of the Gamma-matched
approximation, are derived under the adopted single-reference-port
correlation model. For a different correlation structure, the Gamma
approximation may retain the same diversity-matched form, but its
asymptotic coefficient must be derived from the corresponding joint
distribution and is not generally given by \eqref{eq:a0_final}.}

\textcolor{black}{The index set $\mathcal K_r$ denotes a subset of $r=\operatorname{rank}(\mathbf R)$ port indices that spans the effective spatial signal subspace induced by the correlation matrix $\mathbf R$. In practice, $\mathcal K_r$ can be obtained by selecting any $r$ linearly independent columns of $\mathbf R$, or equivalently any set of $r$ ports for which the corresponding principal submatrix $\mathbf R_{\mathcal K_r,\mathcal K_r}$ is nonsingular. For a full-rank correlation matrix, one may simply set $\mathcal K_r=\{1,\ldots,N\}$. For a rank-deficient or numerically ill-conditioned matrix, $\mathcal K_r$ may be determined using standard rank-revealing procedures such as pivoted Cholesky decomposition, QR decomposition with column pivoting, or eigenvalue-thresholding followed by the selection of the dominant independent ports. Thus, $\mathcal K_r$ is not an additional physical design parameter, but a computational representation of the effective number of independent spatial modes supported by the finite aperture.}

\textcolor{black}{\emph{Remark 3 (Practical construction of $\mathcal K_r$):}
For numerical implementation, the rank $r$ can be determined as the number of eigenvalues of $\mathbf R$ larger than a prescribed tolerance, e.g.,
\begin{equation}
	r=\#\{i:\lambda_i(\mathbf R)>\bar{\epsilon}\lambda_{\max}(\mathbf R)\},
\end{equation}
 \textcolor{black}{where $\#\{\cdot\}$ denotes the cardinality of the enclosed set, i.e.,
	the number of eigenvalues of $\mathbf R$ that exceed
	$\bar{\epsilon}\lambda_{\max}(\mathbf R)$, and $\bar{\epsilon}$ is a
	prescribed relative numerical threshold.} After determining $r$, the set $\mathcal K_r$ can be obtained from a pivoted Cholesky or QR factorization of $\mathbf R$. The selected pivot indices form $\mathcal K_r$. This procedure automatically reduces to $\mathcal K_r=\{1,\ldots,N\}$ when $\mathbf R$ is full rank and all ports provide non-negligible independent spatial observations.}

While \eqref{eq:asymp_outage1} describes the generic asymptotic form of the outage
probability, the previous result identifies the corresponding exponent
and array-gain constant for the considered FAMA system.

\section{Numerical Results}\label{Numerical}
In this section, numerical results are presented to validate the analytical
expressions derived in Sections~\ref{statistical}-\ref{fama} and to illustrate the impact of key
system parameters on the outage performance of FAMA under correlated
Nakagami-$m$ fading. Unless otherwise stated, the desired and interfering
channels are assumed to experience Nakagami fading with parameters
$m_s$ and $m_I$, respectively, while the spatial correlation across the
fluid-antenna ports is generated according to the model in
\eqref{eq:mu_k_uniform}. All users are assumed statistically identical, and
the system operates in an interference-limited regime. The simulation
parameters are summarized in Table~\ref{tab:sim_params}. \textcolor{black}{In all rank-deficient correlation scenarios, the effective rank $r$ and the corresponding index set $\mathcal K_r$ are obtained from the eigenvalue spectrum of $\mathbf R$ using a relative threshold $10^{-6}$, followed by a rank-revealing QR factorization to identify the independent port indices.}
MC simulations are performed to corroborate the analytical outage
probability, its asymptotic approximation, and the Gamma-matched
closed-form expression. For benchmarking purposes, a conventional maximum-ratio combining (MRC) receiver with independent, uncorrelated antennas is considered. \textcolor{black}{We also consider additional FAMA-specific benchmark schemes. First, we include an independent-port FAMA benchmark, where the ports are assumed statistically independent, i.e., $\mathbf R=\mathbf I_N$. This benchmark corresponds to the idealized assumption adopted in several early FAMA studies and provides an optimistic reference for the maximum selection diversity achievable with $N$ ports. Second, we consider a random-port selection scheme, where the receiver selects one of the available ports without using instantaneous SIR information. This benchmark quantifies the gain obtained from opportunistic port selection. Third, we consider a desired-signal-based selection scheme, where the selected port is the one maximizing the desired-signal power $S_k=|g_k|^2$, rather than the instantaneous SIR. This comparison highlights the importance of interference-aware selection in FAMA systems operating in interference-limited environments.}

The results highlight the roles of the number of
ports, fading severity, and spatial correlation rank in determining both the
outage probability and the effective diversity order of the FAMA receiver.
\begin{table}[t]
	\centering
	\caption{Simulation Parameters for Numerical Results}
	\label{tab:sim_params}
	\begin{tabular}{l c}
		\hline
		\hline
		\textbf{Parameter} & \textbf{Value} \\
		\hline
		Number of fluid-antenna ports & $N = 8$ \\[2pt]
		Fluid-antenna aperture width & $W = 5\lambda$ \\[2pt]
		Desired-signal Nakagami parameter & $m_s \in 2$ \\[2pt]
		Interference Nakagami parameter & $m_I \in 2$ \\[2pt]
		Number of interferers & $N_I = 5$ \\[2pt]
		Average signal and interference powers & $\Omega_s = \Omega_I = 1$ \\[2pt]
				\hline
		\hline
	\end{tabular}
\end{table}

\begin{figure}%
	\centering
	\includegraphics[width=0.95\linewidth]{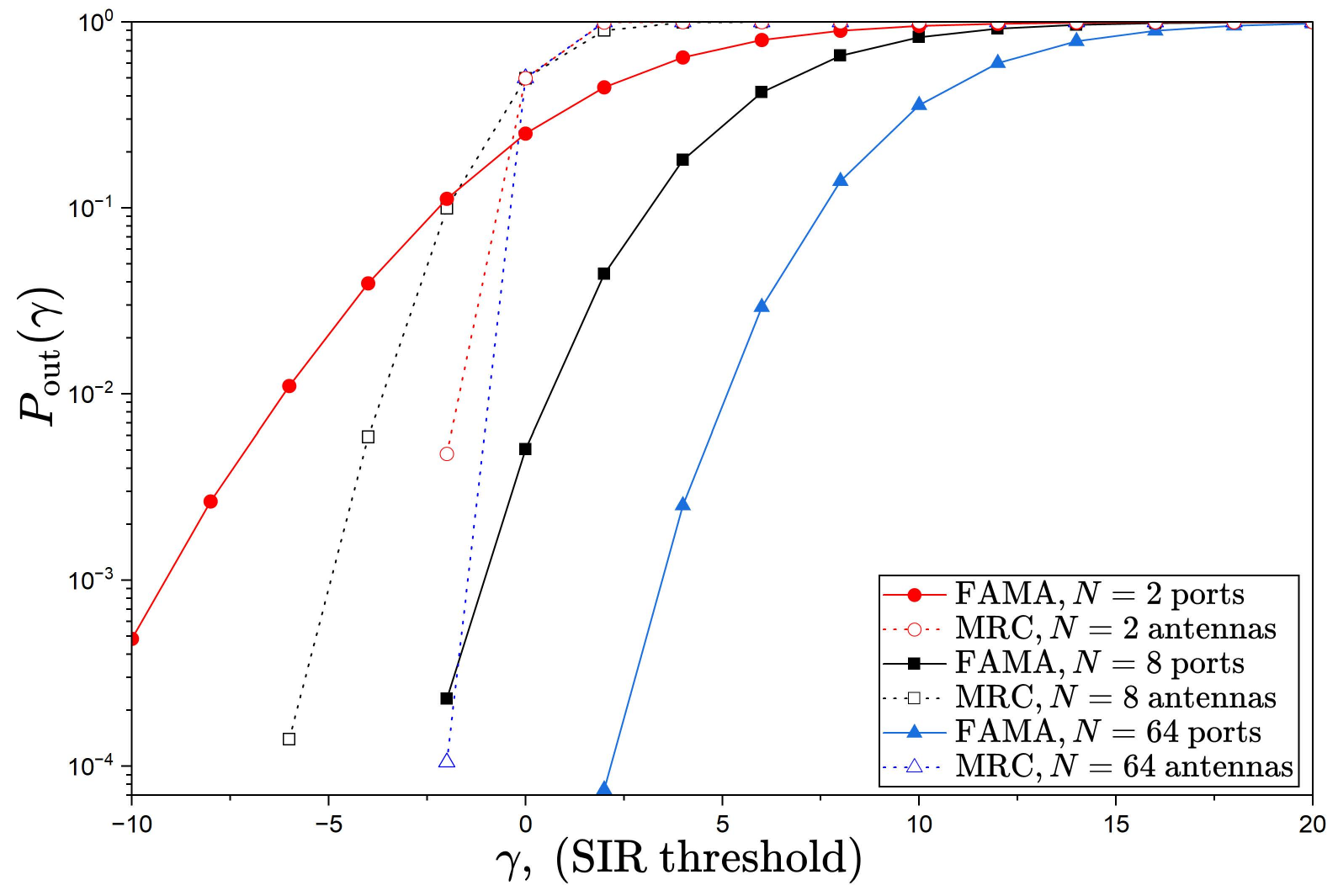}
	\caption{Outage probability versus the SIR threshold for the proposed FAMA receiver and conventional MRC under Nakagami-$m$ fading.}
	\label{fig6}
\end{figure}
Fig.~\ref{fig6} depicts the outage probability $P_{\mathrm{out}}(\gamma)$ of the proposed FAMA receiver as a function of the SIR threshold~$\gamma$ for different numbers of available antenna ports. The performance of FAMA is compared with that of conventional MRC under identical fading and interference conditions. In the FAMA scheme, the receiver selects the antenna port that yields the maximum instantaneous SIR, whereas MRC coherently combines all available signal components without explicitly accounting for interference conditions. 
As the number of ports increases, the FAMA outage curves exhibit a pronounced {leftward} shift, indicating that a significantly lower SIR threshold is required to achieve a given outage probability. This behavior highlights the strong selection gain offered by FAMA, which becomes increasingly effective as the spatial dimensionality grows and the likelihood of encountering favorable interference conditions increases. In contrast, although MRC benefits from array gain as the number of antennas increases, its performance improvement saturates in the interference-limited regime. This is because MRC coherently amplifies both the desired signal and the interference, thereby limiting the achievable SIR gains. As a result, increasing the number of antennas does not necessarily translate into a lower outage probability for MRC and may even lead to a slight rightward shift of the outage curve, as observed for $N=8$ compared to $N=2$.

\textcolor{black}{For comparison purposes, we also define two alternative port-selection strategies. In random-port selection, the selected port index $k_{\mathrm{rand}}$ is uniformly drawn from the set $\{1,\ldots,N\}$, and the corresponding SIR is given by
\begin{equation}
	\mathrm{SIR}_{\mathrm{rand}}=\mathrm{SIR}_{k_{\mathrm{rand}}}.
\end{equation}
In desired-signal-based selection, the receiver selects the port with the strongest desired-signal power, namely
\begin{equation}
	k_{\mathrm{sig}}=\arg\max_{k=1,\ldots,N} |g_k|^2,
\end{equation}
and the resulting SIR is
\begin{equation}
	\mathrm{SIR}_{\mathrm{sig}}=
	\frac{|g_{k_{\mathrm{sig}}}|^2}{|g_{I,k_{\mathrm{sig}}}|^2}.
\end{equation}
These two schemes are used only as numerical benchmarks.}

	\begin{figure}%
	\centering
	\includegraphics[width=0.95\linewidth]{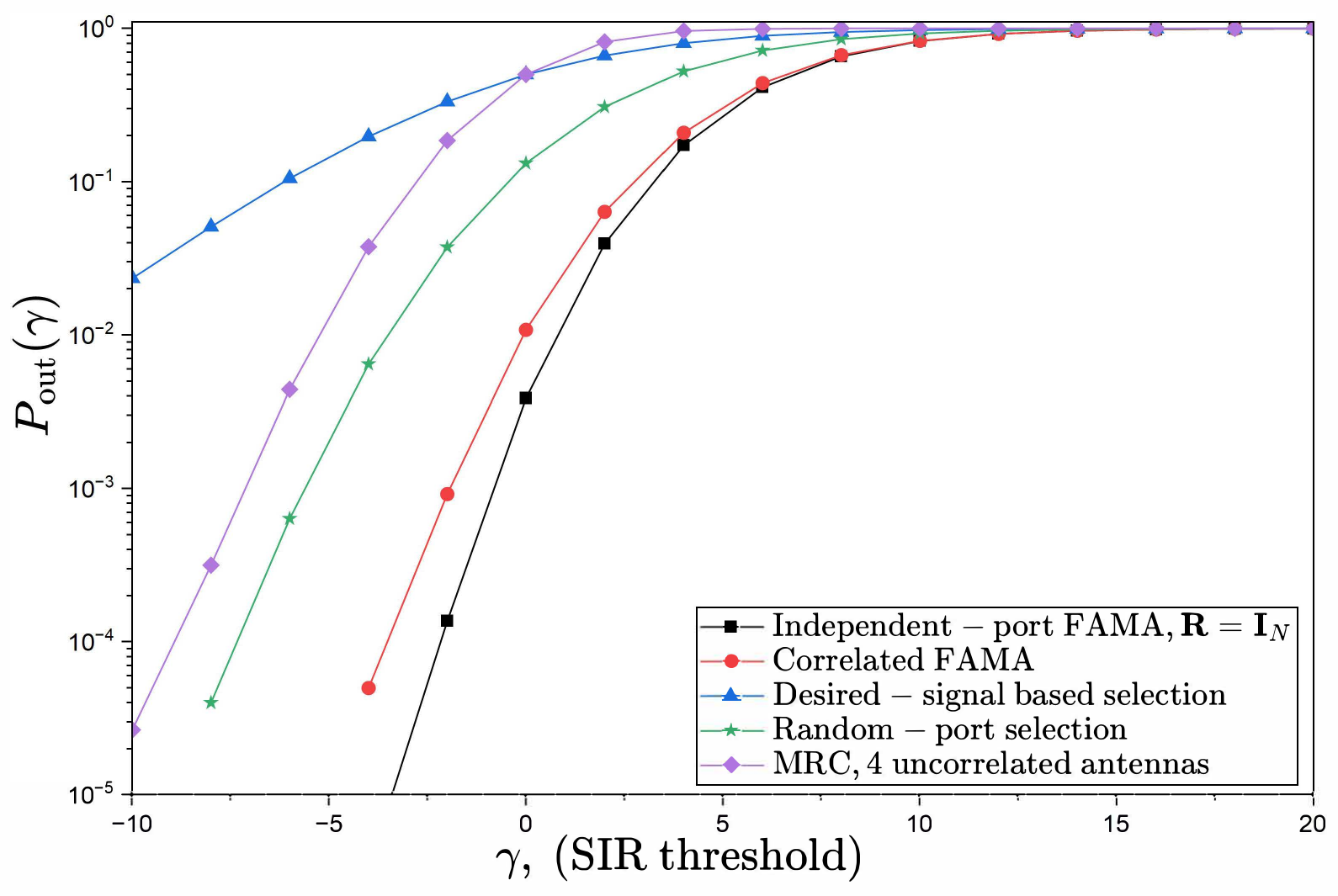}
	\caption{\textcolor{black}{Outage probability versus SIR threshold for correlated FAMA, independent-port FAMA, random-port selection, desired-signal-based selection, and conventional MRC under Nakagami-$m$ fading.}}
	\label{fig10}
\end{figure}

\textcolor{black}{Fig.~\ref{fig10} compares the proposed correlated FAMA scheme with independent-port FAMA, random-port selection, desired-signal-based port selection, and conventional MRC. The independent-port FAMA curve provides an optimistic lower bound, since all ports contribute independent spatial diversity, whereas the correlated FAMA curve is shifted to the right due to finite-aperture spatial correlation. Among the alternative FAMA benchmarks, random-port selection gives the poorest performance because it does not exploit instantaneous channel or interference variations. Desired-signal-based selection improves over random selection by choosing the strongest desired-signal port, but it remains inferior to SIR-based FAMA because it ignores the spatial variation of the aggregate interference. Thus, in interference-limited FAMA systems, the best port is not necessarily the one with the strongest desired signal, but the one that offers the most favorable desired-signal-to-interference balance. These results confirm the importance of both correlation-aware modeling and interference-aware port selection.}

	\begin{figure}%
	\centering
	\includegraphics[width=0.95\linewidth]{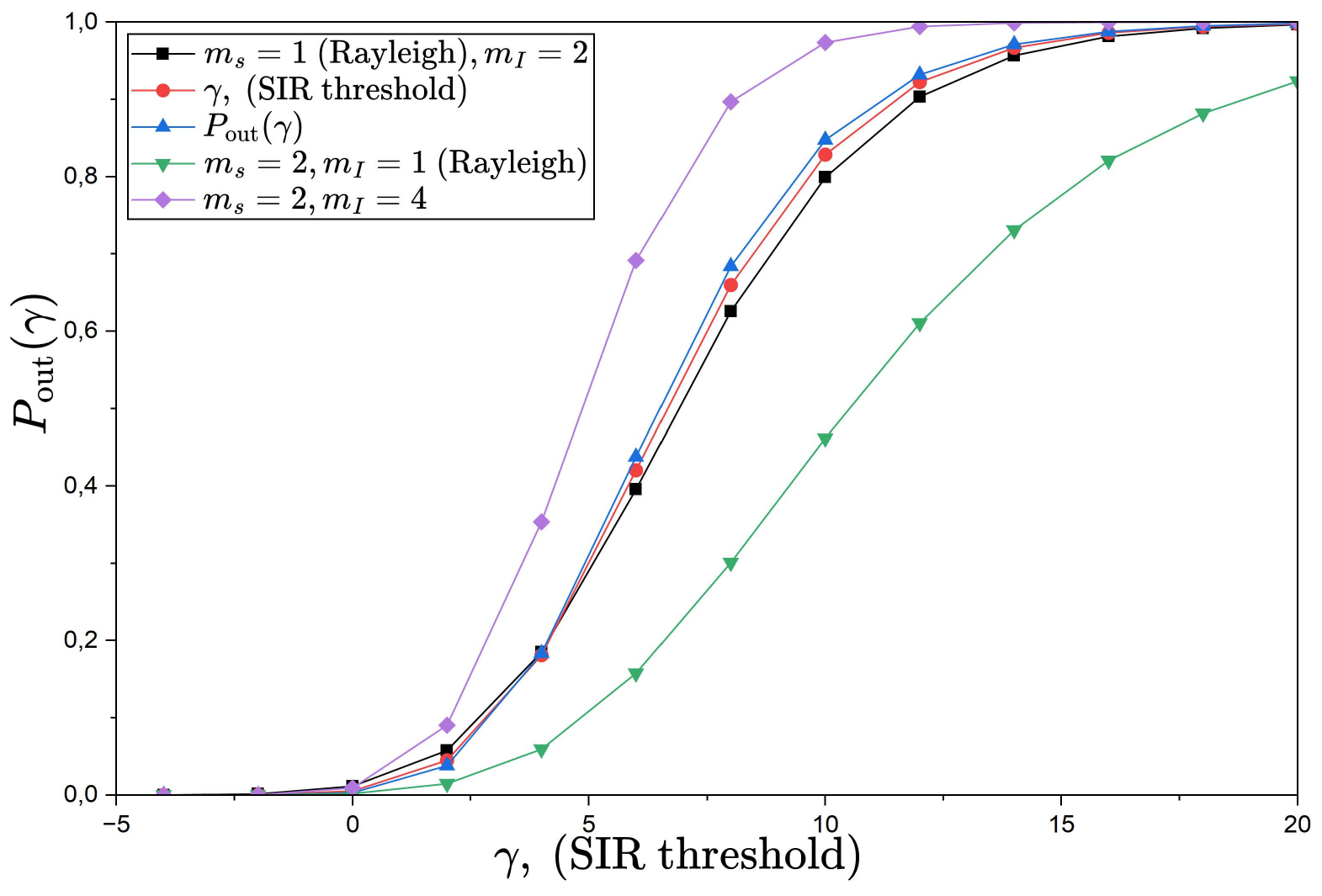}
	\caption{Outage probability versus SIR threshold for different Nakagami fading parameters of the desired and interfering links.}
	\label{fig1}
\end{figure}
Fig.~\ref{fig1} illustrates the outage probability as a function of the SIR threshold for different combinations of the Nakagami fading parameters of the desired and interfering links when $N=8$ ports. Increasing the fading parameter of the desired signal, $m_s$, yields a consistent but relatively moderate reduction in the outage probability, which can be attributed to the decreased fading severity of the useful link while its average power remains unchanged. In contrast, variations in the interference fading parameter $m_I$ have a markedly stronger impact on performance. Specifically, smaller values of $m_I$, corresponding to more severe fading on the interfering links, significantly improve the outage performance by increasing the likelihood of deep interference fades, whereas larger values of $m_I$ lead to a noticeable degradation due to more stable and persistently strong interference. Consequently, the separation between curves associated with different $m_I$ values is substantially larger than that observed when varying $m_s$, particularly in the medium- and high-SIR regimes.These observations indicate that, over the considered SIR range, the outage performance is mainly influenced by the statistics of the interfering signals, whereas the fading severity of the desired link affects the asymptotic decay rate of the outage probability through the effective diversity order, as formally established in Theorem~\ref{Theorem2}.

	\begin{figure}%
	\centering
	\includegraphics[width=0.95\linewidth]{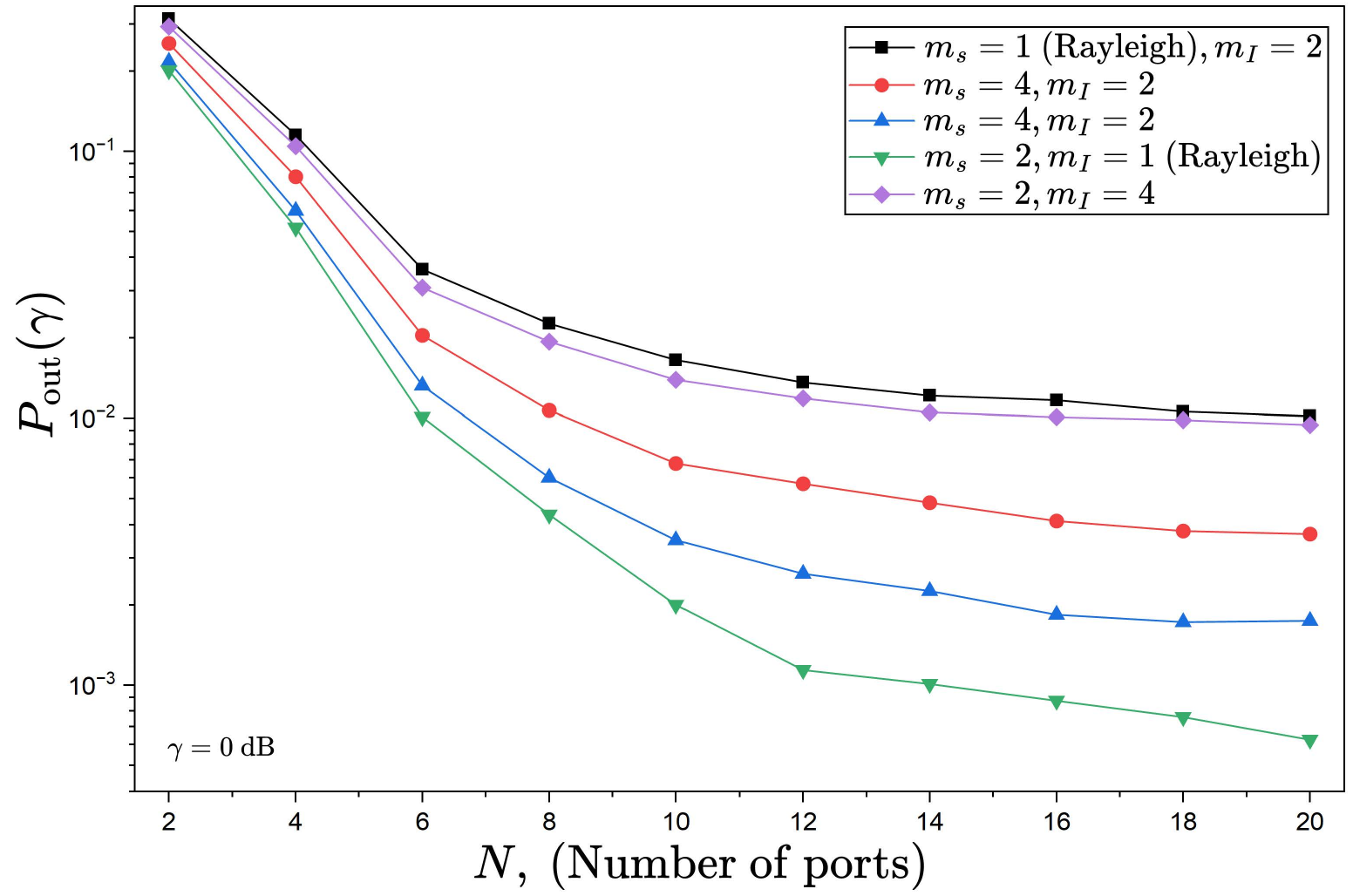}
	\caption{Outage probability versus number of ports for different Nakagami fading parameters of the desired and interfering links.}
	\label{fig2}
\end{figure}
Fig.~\ref{fig2} depicts the outage probability as a function of the number of FAMA ports $N$ for different combinations of the desired- and interfering-channel fading parameters $(m_s,m_I)$. As expected, increasing $N$ leads to a monotonic reduction in the outage probability for all considered cases, reflecting the additional selection diversity offered by a larger number of fluid-antenna ports. For a fixed interference severity $m_I=2$, increasing the desired-channel fading parameter $m_s$ from Rayleigh fading ($m_s=1$) to milder fading conditions ($m_s=2$ and $m_s=4$) results in a consistent but moderate improvement in outage performance. This behavior is in agreement with the diversity analysis in Theorem~\ref{Theorem2}, since $m_s$ affects the effective diversity order and thus primarily influences the asymptotic decay of the outage probability at very low SIR thresholds. In contrast, variations in the interference fading parameter $m_I$ induce a substantially larger shift of the outage curves across the entire range of $N$, highlighting the dominant role of the interference statistics in determining the finite-SIR outage behavior of the considered interference-limited system.

\begin{figure}%
	\centering
	\includegraphics[width=0.95\linewidth]{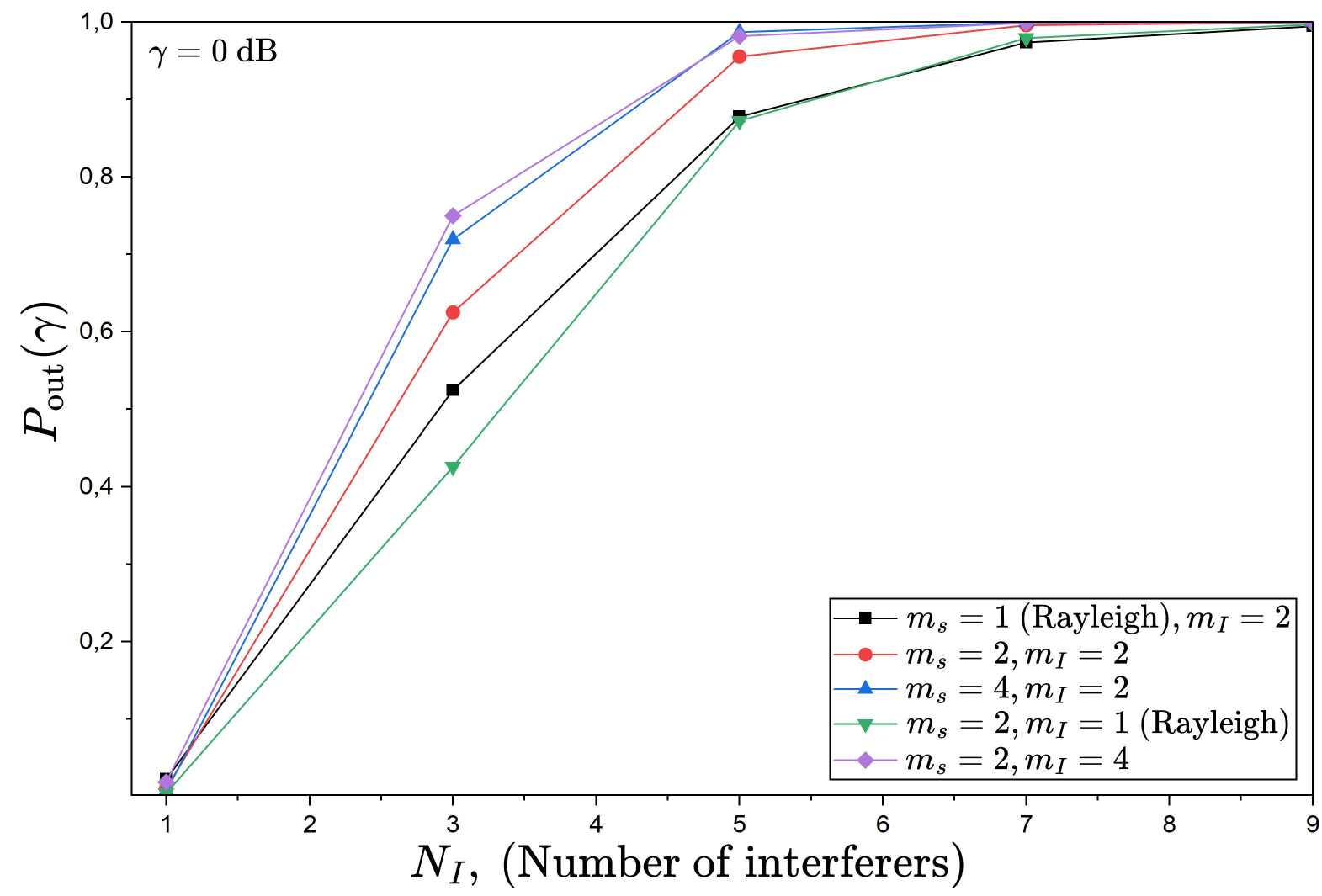}
	\caption{Outage probability versus number of interferers for different Nakagami fading parameters of the desired and interfering links.}
	\label{fig3}
\end{figure}
Fig.~\ref{fig3} depicts the outage probability as a function of the number of interferers $N_I$ for different combinations of the desired- and interfering-channel fading parameters $(m_s,m_I)$, evaluated at a fixed SIR threshold $\gamma=0$~dB. As expected, the outage probability increases monotonically with $N_I$ for all considered scenarios, rapidly approaching unity as the aggregate interference power grows, which reflects the interference-limited nature of the system. For a fixed desired-channel fading parameter $m_s=2$, variations in the interference fading severity $m_I$ result in a pronounced separation among the outage curves, indicating that the statistics of the interfering signals play a dominant role in shaping the overall outage behavior. In contrast, for a fixed interference condition $m_I=2$, increasing the desired-channel fading parameter from Rayleigh fading ($m_s=1$) to milder fading conditions ($m_s=2$ and $m_s=4$) yields a consistent but comparatively moderate performance improvement. This observation is fully consistent with the diversity analysis in Theorem~\ref{Theorem2}, since $m_s$ primarily affects the effective diversity order and thus governs the asymptotic decay of the outage probability at very low outage levels, whereas the present figure corresponds to a fixed SIR threshold and moderate-to-high outage probabilities. 
\textcolor{black}{The result in Fig.~\ref{fig3} can also be interpreted from a multiplexing
	perspective. For a target outage probability $\epsilon$, the
	largest value of $N_I$ for which the outage curve remains below
	$\epsilon$ determines the number of cochannel interferers that
	can be supported while satisfying the required reliability level.
	Therefore, the proposed outage analysis provides an
	outage-constrained measure of the multiplexing capability of
	the FAMA network. For example, if the target reliability is
	specified by $P_{\mathrm{out}}\leq \epsilon$, the admissible number
	of simultaneously active users sharing the same time--frequency
	resource is obtained as $N_I^{\star}+1$, where
	$N_I^{\star}$ is defined by the largest number of interferers
	satisfying the outage constraint. The figure shows that this
	multiplexing capability is strongly affected by the interference
	fading severity $m_I$ and by the desired-link fading parameter
	$m_s$. Moreover, when the ports are spatially correlated, the
	multiplexing gain is ultimately limited by the effective rank of
	the correlation matrix, since only the effective independent
	spatial modes can be used to avoid interference.}

\begin{figure}%
	\centering
	\includegraphics[width=0.95\linewidth]{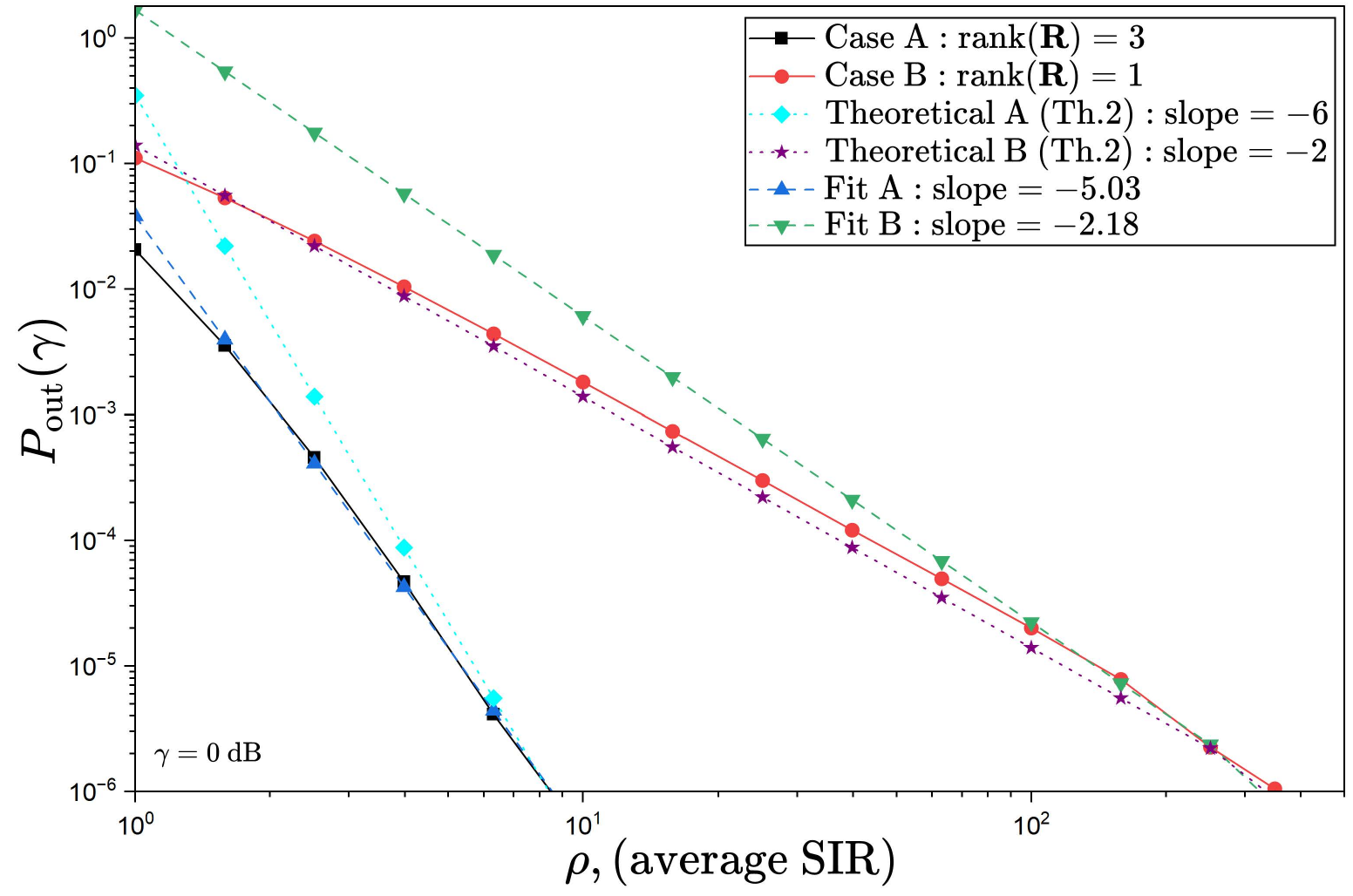}
	\caption{Outage probability  versus average SIR scaling factor. MC results are shown for two correlation scenarios with $\mathrm{rank}(\mathbf{R})=3$ (Case~A) and $\mathrm{rank}(\mathbf{R})=1$ (Case~B).}
	\label{fig4}
\end{figure}
Fig.~\ref{fig4} verifies the diversity order of the proposed FAMA scheme by plotting the outage probability $P_{\mathrm{out}}$ versus the average SIR scaling factor $\rho=\bar{\mathrm{SIR}}$ on a log--log scale. The threshold is fixed to $\gamma=0$~dB, while the average SIR scaling factor scales the average desired-signal power relative to the interference and thus drives the system into the high-SIR regime. Two correlation configurations are examined, generated to have different numerical ranks, namely $\mathrm{rank}(\mathbf{R})=3$ (Case~A) and $\mathrm{rank}(\mathbf{R})=1$ (Case~B), which lead to the theoretical diversity orders $d_{\mathrm{eff}}=m_s\mathrm{rank}(\mathbf{R})=6$ and $d_{\mathrm{eff}}=2$, respectively, for $m_s=2$. MC results are obtained from $2\times10^{7}$ realizations, where the instantaneous SIR is computed per port and the maximum across ports is selected according to the FAMA rule. To quantify the diversity order, an automatic tail-fitting routine selects the last outage samples above the MC reliability floor (approximately $3/N_{\mathrm{sims}}$) and performs a linear regression of $\log_{10} P_{\mathrm{out}}$ versus $\log_{10}\rho$, yielding slopes of about $-5.03$ (Case~A) and $-2.18$ (Case~B). Moreover, the figure superimposes the asymptotic Theorem~\ref{Theorem2} trends, i.e., $P_{\mathrm{out}}\propto \rho^{-d_{\mathrm{eff}}}$, anchored at the high-SIR tail.

\begin{figure}%
	\centering
	\includegraphics[width=0.95\linewidth]{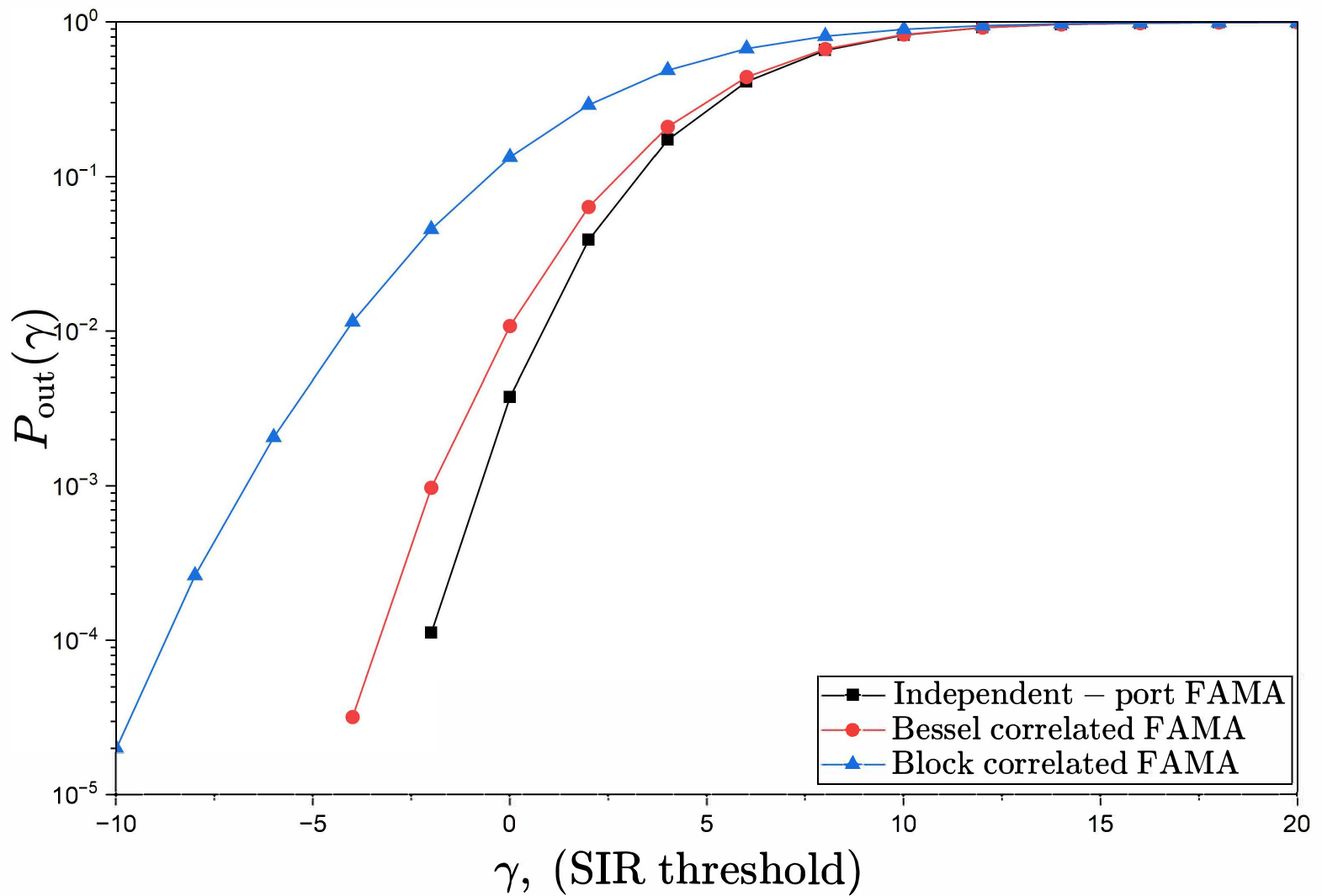}
	\caption{\textcolor{black}{Outage probability versus SIR threshold for independent-port FAMA,
			classical Bessel-correlated FAMA, and block-correlated FAMA under
			Nakagami-$m$ fading.}}
	\label{fig11}
\end{figure}

\textcolor{black}{To evaluate the impact of spatial correlation modeling, we compare three FAMA benchmarks: independent-port FAMA, Bessel-correlated FAMA, and block-correlated FAMA based on \cite{Ramirez2024BlockCorrelation}. The independent-port case represents the idealized assumption that all ports provide independent fading observations and therefore gives the most optimistic performance. The Bessel-correlated model accounts for finite-aperture spatial correlation through the classical isotropic-scattering correlation function and serves as a tractable analytical baseline. The block-correlated model partitions the $N$ ports into spatial coherence blocks with constant intra-block correlation, capturing the diversity saturation caused by dense port sampling within a finite aperture.}

\textcolor{black}{Fig.~\ref{fig11} shows that the block-correlated benchmark yields the highest outage probability because ports within the same coherence block provide less independent selection diversity. This confirms that simplified or independent-port assumptions can overestimate the reliability gain of FAMA, while the qualitative conclusion remains unchanged: the achievable gain is governed by the effective spatial degrees of freedom supported by the aperture rather than by the nominal number of ports alone.}

\begin{figure}%
	\centering
	\includegraphics[width=0.95\linewidth]{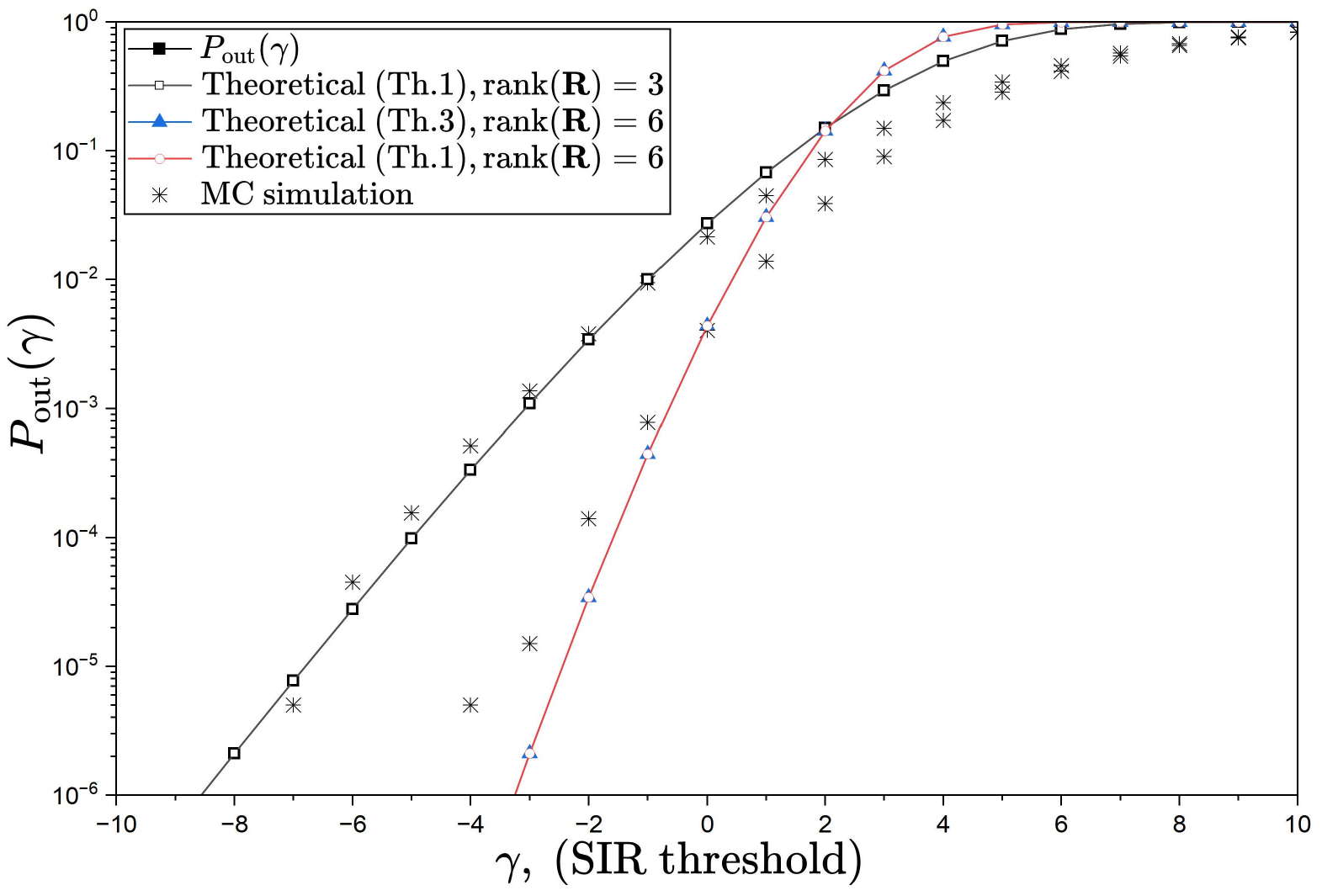}
	\caption{Outage probability in terms of exact analysis (Theorem 1) and Gamma-matched approximation (Theorem 3) versus the SIR threshold. MC results are shown for two correlation scenarios with $\mathrm{rank}(\mathbf{R})=3$ (Case~A) and $\mathrm{rank}(\mathbf{R})=1$ (Case~B).}
	\label{fig5}
\end{figure}
Fig.~\ref{fig5} validates the Gamma-matched outage approximation proposed in Theorem~\ref{Theorem3} by comparing it with the exact analytical expression of Theorem~\ref{Theorem1} and MC simulations. The outage probability $P_{\mathrm{out}}(\gamma)$ is depicted as a function of the SIR threshold $\gamma$. Two correlation scenarios are considered, corresponding to correlation matrices of ranks $\mathrm{rank}(\mathbf{R})=3$ and $\mathrm{rank}(\mathbf{R})=6$, which yield different effective diversity orders $m_s\,\mathrm{rank}(\mathbf{R})$ according to Theorem~3.
The figure demonstrates an excellent agreement between the Gamma-matched approximation and the exact analytical outage probability across the entire range of practical SIR thresholds. As the rank of the correlation matrix increases, the outage curve exhibits a sharper transition, reflecting the higher effective diversity order predicted by Theorem~\ref{Theorem2}. Importantly, the Gamma-matched approximation remains accurate not only in the high-SIR asymptotic regime but also in the moderate-outage region, thereby justifying its use as a low-complexity yet reliable alternative to the exact analysis. 

\textcolor{black}{The above observations lead to several practical design guidelines. First, increasing the number of available ports is beneficial only when the additional ports correspond to sufficiently distinct spatial channel observations. If the aperture is too compact, the correlation matrix becomes rank deficient or ill-conditioned, and the diversity gain saturates despite increasing $N$. Therefore, port densification should be accompanied by a sufficiently large effective aperture. Otherwise, it mainly increases switching and probing complexity without providing proportional outage reduction. Second, the results demonstrate that the desired-link fading parameter $m_s$ and the interference-link fading parameter $m_I$ play different roles. While $m_s$ determines the asymptotic diversity slope, $m_I$ has a strong impact on the finite-SIR outage level because severe interference fading increases the probability of finding ports with deep interference nulls. This suggests that FAMA is particularly attractive in interference-limited networks where the interference field exhibits strong spatial fluctuations across the fluid-antenna aperture. Third, the proposed Gamma-matched approximation can be used as a low-complexity design tool to estimate the number of ports or the aperture size required to meet a target outage probability without resorting to exhaustive Monte Carlo simulations.}
\section{Conclusion}
\label{conclusion}

This paper presented an analytical framework for characterizing the outage performance of FAMA systems over spatially correlated Nakagami-$m$ fading channels. \textcolor{black}{By accounting for the joint statistics of the desired
and interfering signals across the fluid-antenna ports, we first
presented a general joint-distribution-based outage formulation and then
derived an explicit outage expression under the adopted
single-reference-port correlated Nakagami-$m$ model.} To enable tractable evaluation beyond the asymptotic regime, we introduced a Gamma-matched outage approximation that preserves both the diversity order and the array-gain constant. The resulting closed-form expression accurately approximates the outage probability over a wide range of practical SIR thresholds while significantly reducing computational complexity.

\textcolor{black}{In addition to reliability characterization, the outage analysis provides an outage-constrained interpretation of the multiplexing capability of FAMA networks. For a given SIR threshold and target outage probability, the maximum number of simultaneously supported cochannel users is determined by the largest number of interferers for which the outage probability remains below the target constraint. This shows that FAMA multiplexing gain increases with the number of effective independent spatial modes, but is reduced by spatial correlation and interference conditions that limit deep interference fades.}

\textcolor{black}{From a design perspective, the results show that FAMA performance is governed by the effective spatial degrees of freedom created by the aperture, rather than by the nominal number of ports. They also show that Nakagami-$m$ fading offers engineering insight beyond Rayleigh-only models, since $m_s$ determines the outage diversity slope, whereas $m_I$ mainly controls the finite-SIR outage level. These findings guide port-density selection, aperture sizing, and the identification of interference-limited scenarios where FAMA can provide substantial reliability gains.}

Extensive MC simulations corroborated the exact analysis, the asymptotic characterization, and the accuracy of the proposed approximation under various fading and correlation conditions.

\begin{appendices}
	
	\renewcommand{\theequation}{\Alph{section}.\arabic{equation}}
	\makeatletter
	\@addtoreset{equation}{section}
	\makeatother
\appendices

\section{Proof of Theorem~\ref{Theorem1}}
\label{Theorem1proof}

Let $X_k \triangleq |h_k|^2$ and $Y_k \triangleq |h_{I,k}|^2$ denote the
instantaneous desired-signal and aggregate-interference powers at port
$k$, respectively. The instantaneous SIR at port $k$ is
\begin{equation}
	\mathrm{SIR}_k = \frac{X_k}{Y_k}.
\end{equation}
The SIR outage probability is
\begin{equation}
	P_{\mathrm{out}}(\gamma)
	=
	\Pr\!\left(
	X_1 < \gamma Y_1,\ldots,X_N < \gamma Y_N
	\right).
	\label{eq:outage_def}
\end{equation}

Conditioning on the interference power vector
$\mathbf{Y}=(Y_1,\ldots,Y_N)$ and applying the law of total probability,
we obtain
	\begin{align}
		P_{\mathrm{out}}(\gamma)
		&\!=\!
		\int_{0}^{\infty}\!\!\!\cdots\!\int_{0}^{\infty}
		\Pr\!\!\left(
		X_1 < \gamma y_1,\ldots,X_N < \gamma y_N
		\,\big|\, \mathbf{Y}=\mathbf{y}
		\right)
		\nn\\
		&\hspace{2.0cm}\times
		f_{\mathbf{Y}}(y_1,\ldots,y_N)\,
		\mathrm{d}y_1\cdots\mathrm{d}y_N .	\label{eq:conditioning_on_Y}
	\end{align}

For a fixed realization $\mathbf{Y}=\mathbf{y}$, the joint probability
term in \eqref{eq:conditioning_on_Y} is simply the joint CDF of the
desired-signal powers evaluated at $(\gamma y_1,\ldots,\gamma y_N)$.
Using the reference-port decomposition, this CDF can be written as
\begin{align}
	&\Pr\!\left(
	X_1 < \gamma y_1,\ldots,X_N < \gamma y_N
	\,\big|\, \mathbf{Y}=\mathbf{y}
	\right)\nn\\
	&\quad\quad\quad\quad=
	\int_{0}^{\gamma y_1}
	f_{X_1}(t)
	\prod_{k=2}^{N}
	F_{X_k|X_1}(\gamma y_k \mid t)\, dt,
	\label{eq:joint_cdf_step}
\end{align}
where $f_{X_1}(t)$ is the marginal PDF of $X_1$ and
$F_{X_k|X_1}(\cdot|\cdot)$ denotes the conditional CDF of $X_k$ given $X_1$.

The desired-signal powers $\{X_k\}$ follow a spatially correlated
Nakagami-$m_s$ model. Thus, the marginal PDF of $X_1$ is Gamma distributed as
\begin{equation}
	f_{X_1}(t)
	=
	\frac{m_s^{m_s}}{\Gamma(m_s)}
	t^{m_s-1} e^{-m_s t}.
	\label{eq:pdf_X1}
\end{equation}
From Theorem~1 in \cite{Tlebaldiyeva2022}, the joint CDF of
correlated Nakagami-$m_s$ envelopes admits a product form in which
each factor corresponds to the conditional CDF of $X_k$ given $X_1$.
Therefore, conditioning on $X_1=t$, the conditional CDF of $X_k$
can be directly read as
\begin{equation}
	F_{X_k|X_1}(\gamma y \mid t)
	=
	1-
	Q_{m_s}\!\left(
	\sqrt{\frac{2m_s\mu_k^2}{1-\mu_k^2}\,t},
	\sqrt{\frac{2m_s\gamma}{1-\mu_k^2}\,y}
	\right).
	\label{eq:conditional_cdf_signal}
\end{equation}

We now average over the interference powers. Conditioning on the
reference pair $(X_1,Y_1)=(t,z)$, define for $k\ge 2$
\begin{equation}
	\Psi_k(t,z;\gamma)
	\triangleq
	\Pr\!\left(X_k<\gamma Y_k \,\big|\, X_1=t,\,Y_1=z\right).
	\label{eq:Psi_def}
\end{equation}
Applying the law of total probability with respect to $Y_k$ yields
	\begin{align}
		\Psi_k(t,z;\gamma)
		&=
		\int_{0}^{\infty}
		\Pr\!\left(
		X_k<\gamma y
		\,\big|\, X_1=t,\,Y_k=y,\,Y_1=z
		\right)\nn\\
		&\times
		f_{Y_k|Y_1}(y|z)\, dy.	\label{eq:Psi_total_probability}
	\end{align}

Since the desired-signal and interference processes are independent,
conditioning on $(Y_k,Y_1)$ does not affect the distribution of $X_k$
given $X_1=t$. Therefore, we have
\begin{equation}
	\Pr\!\left(
	X_k<\gamma y
	\,\big|\, X_1=t,\,Y_k=y,\,Y_1=z
	\right)
	=
	F_{X_k|X_1}(\gamma y \mid t).
	\label{eq:independence_argument}
\end{equation}
Substituting \eqref{eq:independence_argument} into
\eqref{eq:Psi_total_probability} gives
\begin{equation}
	\Psi_k(t,z;\gamma)
	=
	\int_{0}^{\infty}
	F_{X_k|X_1}(\gamma y \mid t)\,
	f_{Y_k|Y_1}(y|z)\, dy.
	\label{eq:Psi_expectation_form}
\end{equation}

The conditional PDF $f_{Y_k \mid Y_1}(y \mid z)$ follows from the
envelope-domain conditional distribution in Proposition~1 of~\cite{Tlebaldiyeva2022}
via a deterministic change of variables, by defining the interference
power gains as
$Y_k = |g_{I,k}|^2$ and $ Y_1 = |g_{I,1}|^2 
$, and is given by
	\begin{align}
		f_{Y_k|Y_1}(y|z)
		&=
		\frac{m_I}{1-\mu_k^2}
		\left(
		\frac{y}{\mu_k^2 z}
		\right)^{\frac{m_I-1}{2}}
		\exp\!\left(
		-\frac{m_I}{1-\mu_k^2}(y+\mu_k^2 z)
		\right)
	\nn	\\
		&\hspace{2cm}\times
		I_{m_I-1}\!\left(
		\frac{2m_I\mu_k}{1-\mu_k^2}\sqrt{yz}
		\right).	\label{eq:conditional_pdf_interference}
	\end{align}

Substituting \eqref{eq:conditional_cdf_signal} and
\eqref{eq:conditional_pdf_interference} into
\eqref{eq:Psi_expectation_form} yields
\begin{equation}
	\Psi_k(t,z;\gamma)
	\!=\!\!\!
	\int_{0}^{\infty}\!
	\left[
	1-
	Q_{m_s}\!\left(
	\sqrt{\alpha_k t},\sqrt{\beta_k\gamma y}
	\right)
	\right]
	f_{Y_k|Y_1}(y|z)\, dy,
	\label{eq:psi_def}
\end{equation}
where $\alpha_k=\frac{2m_s\mu_k^2}{1-\mu_k^2}$ and
$\beta_k=\frac{2m_s}{1-\mu_k^2}$.

Finally, substituting \eqref{eq:psi_def} into
\eqref{eq:joint_cdf_step}, and noting that $Y_1$ follows a Gamma
distribution with PDF
\begin{equation}
	f_{Y_1}(z)
	=
	\frac{m_I^{m_I}}{\Gamma(m_I)}
	z^{m_I-1} e^{-m_I z},\label{pdf1}
\end{equation}
we obtain
	\begin{align}
		P_{\mathrm{out}}(\gamma)
		&=\!\!\!
		\int_{0}^{\infty}\!\!
		f_{Y_1}(z)
		\int_{0}^{\gamma z}
		f_{X_1}(t)
		\prod_{k=2}^{N}
		\Psi_k(t,z;\gamma)
		\, dt\, dz.
	\end{align}
\textcolor{black}{We next derive an explicit representation of
$\Psi_k(t,z;\gamma)$. From \eqref{eq:psi_def}, this quantity is the
conditional probability that the desired-signal power at port $k$ is
smaller than $\gamma$ times the interference power at the same port,
given the reference-port realization $(X_1,Y_1)=(t,z)$. Its integrand
contains the complement of a generalized Marcum-$Q$ function of order
$m_s$. For integer $m_s$, this function satisfies the finite recursion
\cite[Eq.~4.54]{Simon2004}
\begin{equation}
	Q_m(\alpha,\beta)
	=
	\exp\!\left(-\frac{\alpha^2+\beta^2}{2}\right)\!
	\sum_{n=1}^{m-1}\!\!
	\left(\frac{\beta}{\alpha}\right)^{n}\!\!
	I_n(\alpha\beta)
	+
	Q_1(\alpha,\beta).\nn
\end{equation}
where $I_n(\cdot)$ is the modified Bessel function of the first kind and
order $n$. In the present derivation, $\alpha=\sqrt{\alpha_k t}$ and
$\beta=\sqrt{\beta_k\gamma y}$. Consequently,
$1-Q_{m_s}(\alpha,\beta)$ equals the order-one contribution
$1-Q_1(\alpha,\beta)$ minus the finite sum in the above recursion. The
minus sign therefore follows directly from taking the complement of
$Q_{m_s}(\alpha,\beta)$.}

\textcolor{black}{Substituting this recursion and the above arguments into
\eqref{eq:psi_def} yields
	\begin{align}
	&	\Psi_k(t,z;\gamma)
		=
		\int_{0}^{\infty}
		\Bigg[
		1
		-
		Q_1\!\left(
		\sqrt{\alpha_k t},\sqrt{\beta_k\gamma y}
		\right)
		\nn\\
		&
		-
		\exp\!\left(
		-\frac{\alpha_k t+\beta_k\gamma y}{2}
		\right)
		\sum_{n=1}^{m_s-1}
		\left(
		\frac{\beta_k\gamma y}{\alpha_k t}
		\right)^{\frac{n}{2}}
		I_n\!\left(
		\sqrt{\alpha_k\beta_k\gamma t y}
		\right)
		\Bigg]
		\nn\\
		&\qquad\times
		f_{Y_k|Y_1}(y|z)\,dy .	\label{eq:Psi_substituted}
	\end{align}}

\textcolor{black}{The integrand in \eqref{eq:Psi_substituted} consists of the order-one
Marcum-$Q$ contribution and the finite correction associated with
$m_s>1$. Hence, by the linearity of integration,
$\Psi_k(t,z;\gamma)$ can be decomposed as
\begin{equation}
	\Psi_k(t,z;\gamma)
	=
	\Psi_k^{(1)}(t,z;\gamma)
	-
	\Psi_k^{(2)}(t,z;\gamma),
	\label{eq:Psi_split}
\end{equation}
where the two terms are defined as follows:}

\textcolor{black}{\begin{align}
	\Psi_k^{(1)}(t,z;\gamma)
&	\triangleq
	\int_{0}^{\infty}
	\Big[
	1-
	Q_1\!\left(
	\sqrt{\alpha_k t},\sqrt{\beta_k\gamma y}
	\right)
	\Big]
	f_{Y_k|Y_1}(y|z)\,dy .
	\label{eq:Psi_1}\\
		\Psi_k^{(2)}(t,z;\gamma)
	&\triangleq
	\int_{0}^{\infty}
	\exp\!\left(
	-\frac{\alpha_k t+\beta_k\gamma y}{2}
	\right)
	\sum_{n=1}^{m_s-1}
	\left(
	\frac{\beta_k\gamma y}{\alpha_k t}
	\right)^{\frac{n}{2}}
	\nn\\
	&\qquad\times
	I_n\!\left(
	\sqrt{\alpha_k\beta_k\gamma t y}
	\right)
	f_{Y_k|Y_1}(y|z)\,dy .	\label{eq:Psi_2}
\end{align}}

\textcolor{black}{The term $\Psi_k^{(1)}(t,z;\gamma)$ is the contribution associated with
the first-order Marcum-$Q$ function and coincides with the conditional
outage term for Rayleigh desired-signal fading ($m_s=1$). In contrast,
$\Psi_k^{(2)}(t,z;\gamma)$ is the finite correction that converts the
order-one result into the Nakagami-$m_s$ result. Indeed, when $m_s=1$,
the summation from $n=1$ to $m_s-1$ is empty, and hence
$\Psi_k^{(2)}(t,z;\gamma)=0$ and
$\Psi_k(t,z;\gamma)=\Psi_k^{(1)}(t,z;\gamma)$. For $m_s>1$, the
correction must be retained. Therefore, we next evaluate
$\Psi_k^{(2)}(t,z;\gamma)$ explicitly.}

Substituting the conditional PDF $f_{Y_k|Y_1}(y|z)$ into
\eqref{eq:Psi_2} gives
	\begin{align}
		&\Psi_k^{(2)}(t,z;\gamma)
		=
		\sum_{n=1}^{m_s-1}
		\frac{e^{-\frac{\alpha_k t}{2}}}{(\alpha_k t)^{\frac{n}{2}}}
		(\beta_k\gamma)^{\frac{n}{2}}
		\nn\\
		&\quad\times
		\int_{0}^{\infty}
		y^{\frac{m_I+n-1}{2}}
		\exp\!\left(
		-\left(
		\frac{\beta_k\gamma}{2}
		+
		\frac{m_I}{1-\mu_k^2}
		\right)y
		\right)
	\nn	\\
		&\quad\times
		I_n\!\left(
		\sqrt{\alpha_k\beta_k\gamma t y}
		\right)
		I_{m_I-1}\!\left(
		\frac{2m_I\mu_k}{1-\mu_k^2}\sqrt{yz}
		\right)
		\,dy .	\label{eq:Psi_double_bessel}
	\end{align}

For the sake of simplicity, we denote	
		\begin{align}
			&\Psi_k^{(2)}(t,z;\gamma)
			=
			\sum_{n=1}^{m_s-1}
			\frac{e^{-\frac{\alpha_k t}{2}}}{(\alpha_k t)^{\frac{n}{2}}}
			(\beta_k\gamma)^{\frac{n}{2}}
		\nn	\\
			&\!\!\times\!\!
			\int_{0}^{\infty}\!\!
			y^{\frac{m_I+n-1}{2}}
			\exp(-c_k y)
			I_n\!\left(
			\sqrt{\alpha_k\beta_k\gamma t y}
			\right)
			I_{m_I-1}\!\left(
			B_k\sqrt{y}
			\right)
			\,dy ,		\label{eq:psi2_start}
		\end{align}

	where $c_k \triangleq \frac{\beta_k\gamma}{2}+\frac{m_I}{1-\mu_k^2},
	\qquad
	B_k \triangleq \frac{2m_I\mu_k}{1-\mu_k^2}\sqrt{z}$.

	For integer $n\ge 0$, the modified Bessel function of the first kind admits
	the power-series representation \cite[Eq.~(9.6.10)]{Abramowitz1964}
	\begin{equation}
		I_n(x)
		=
		\sum_{r=0}^{\infty}
		\frac{1}{r!\,\Gamma(r+n+1)}
		\left(\frac{x}{2}\right)^{2r+n},
		\label{eq:bessel_series_app}
	\end{equation}
	which converges for all $x\in\mathbb{R}$.
	
	Substituting $x=\sqrt{\alpha_k\beta_k\gamma t y}$ into
	\eqref{eq:bessel_series_app} yields
	\begin{equation}
		I_n\!\left(\sqrt{\alpha_k\beta_k\gamma t y}\right)
		=
		\sum_{r=0}^{\infty}
		\frac{1}{r!\,\Gamma(r+n+1)}
		\left(\frac{\alpha_k\beta_k\gamma t}{4}\right)^{r+\frac{n}{2}}
		y^{r+\frac{n}{2}} .
		\label{eq:bessel_series_sub}
	\end{equation}

Now, inserting \eqref{eq:bessel_series_sub} into \eqref{eq:psi2_start} gives
		\begin{align}
			&\Psi_k^{(2)}(t,z;\gamma)
			\!=\!\!
			\sum_{n=1}^{m_s\!-\!1}\!\!
			\sum_{r=0}^{\infty}
			\frac{e^{-\frac{\alpha_k t}{2}}}{r!\,\Gamma(r+n+1)}
		\!\!	\left(\frac{\beta_k\gamma}{\alpha_k t}\right)^{\!\!\frac{n}{2}}
		\!\!\!	\left(\frac{\alpha_k\beta_k\gamma t}{4}\right)^{\!\!r+\frac{n}{2}}
			\nn\\
			&\quad\times
			\int_{0}^{\infty}
			y^{r+\frac{m_I+2n-1}{2}}
			\exp(-c_k y)
			I_{m_I-1}\!\left(B_k\sqrt{y}\right)
			\,dy .		\label{eq:psi2_after_exchange}
		\end{align}

	The interchange of summation and integration is justified since all terms
	are nonnegative for $y\ge 0$ and $c_k>0$, ensuring absolute convergence.
	
Define $d_k \triangleq \frac{B_k}{2}
= \frac{m_I\mu_k}{1-\mu_k^2}\sqrt{z} $ and $\nu \triangleq r+\frac{m_I+2n-1}{2}.$ Then,  the integral in \eqref{eq:psi2_after_exchange} takes the form
\begin{equation}
	\int_{0}^{\infty}
	y^{\nu}
	e^{-c_k y}
	I_{m_I-1}\!\left(2 d_k \sqrt{y}\right)\,dy.
	\label{eq:single_bessel_form_app}
\end{equation}

The integral in \eqref{eq:single_bessel_form_app} is evaluated by expanding
the modified Bessel function into its power series and performing
termwise integration using the Gamma integral, which yields a closed-form
expression in terms of the confluent hypergeometric function ${}_1F_1(\cdot)$.

Specifically, one obtains
\begin{align}
	&\int_{0}^{\infty}
	y^{\nu}
	e^{-c y}
	I_{\mu}\!\left(2 d\sqrt{y}\right)\,dy
	=
	\frac{\Gamma(\nu+1)}{c^{\nu+1}}
	\left(\frac{d}{c}\right)^{\mu}
	\nn\\
	&\times
	{}_1F_1\!\left(\nu+\mu+1;\mu+1;\frac{d^2}{c}\right),
	\label{eq:single_bessel_closed_form}
\end{align}
valid for $\Re(c)>0$ and $\Re(\nu)>-1$.
Specifically, this integral is evaluated by using the power-series representation of $I_{\mu}(\cdot)$, and substituting $x=2d\sqrt{y}$, we obtain
\begin{equation}
	I_{\mu}(2d\sqrt{y})
	=
	\sum_{k=0}^{\infty}
	\frac{d^{2k+\mu}}{k!\,\Gamma(k+\mu+1)}
	\,y^{k+\frac{\mu}{2}}.
	\label{eq:I_series_sub}
\end{equation}

Substituting \eqref{eq:I_series_sub} into \eqref{eq:single_bessel_form_app} and interchanging
the order of summation and integration (justified by absolute convergence
for $\Re(c)>0$) yields
\begin{equation}
	\mathcal{I}(\nu,\mu;c,d)
	=
	\sum_{k=0}^{\infty}
	\frac{d^{2k+\mu}}{k!\,\Gamma(k+\mu+1)}
	\int_{0}^{\infty}
	y^{\nu+k+\frac{\mu}{2}} e^{-cy}\,dy .\nn
\end{equation}

The inner integral is evaluated using the Gamma integral
\[
\int_{0}^{\infty} y^{a-1} e^{-cy}\,dy
=
\frac{\Gamma(a)}{c^{a}},
\qquad \Re(c)>0,\ \Re(a)>0,
\]
with $a=\nu+k+\frac{\mu}{2}+1$, which gives
\begin{equation}
	\mathcal{I}(\nu,\mu;c,d)
	=
	\frac{d^{\mu}}{c^{\nu+\frac{\mu}{2}+1}}
	\sum_{k=0}^{\infty}
	\frac{\Gamma\!\left(\nu+k+\frac{\mu}{2}+1\right)}
	{\Gamma(k+\mu+1)}
	\frac{1}{k!}
	\left(\frac{d^2}{c}\right)^k .
	\label{eq:I_series_gamma_ratio}
\end{equation}

Using the Pochhammer symbol $(x)_k=\Gamma(x+k)/\Gamma(x)$, the ratio of Gamma
functions can be expressed as
\[
\frac{\Gamma\!\left(\nu+k+\frac{\mu}{2}+1\right)}{\Gamma(k+\mu+1)}
=
\frac{\Gamma\!\left(\nu+\frac{\mu}{2}+1\right)}{\Gamma(\mu+1)}
\frac{\left(\nu+\frac{\mu}{2}+1\right)_k}{(\mu+1)_k}.
\]
Substituting this into \eqref{eq:I_series_gamma_ratio} and rewriting the
Gamma-function ratio using Pochhammer symbols yields a series of the form
$\sum_{r=0}^\infty \frac{(a)_r}{(b)_r}\frac{x^r}{r!}$, which by definition
corresponds to the confluent hypergeometric function
${}_1F_1(a;b;x)$ \cite[Eq.~(13.1.2)]{Abramowitz1964}, leading to \eqref{eq:single_bessel_closed_form}.

Applying \eqref{eq:single_bessel_closed_form} to
\eqref{eq:single_bessel_form_app} yields a closed-form expression for each
$(n,r)$ term in \eqref{eq:psi2_after_exchange}. Consider the $r$-dependent series
\begin{equation}
	S
	\triangleq
	\sum_{r=0}^{\infty}
	\frac{1}{r!}\,
	\frac{\Gamma(r+n+m_I)}{\Gamma(r+n+1)}
	\left(\frac{\alpha_k\beta_k\gamma t}{4c_k}\right)^{r}.
	\label{eq:S_def}
\end{equation}

Using the Pochhammer symbol $(a)_r \triangleq \Gamma(a+r)/\Gamma(a)$, we write
\begin{align}
&	\frac{\Gamma(r+n+m_I)}{\Gamma(r+n+1)}
	=
	\frac{\Gamma(n+m_I)}{\Gamma(n+1)}
	\frac{\Gamma(r+n+m_I)}{\Gamma(n+m_I)}
	\frac{\Gamma(n+1)}{\Gamma(r+n+1)}
	\nn\\
	&=
	\frac{\Gamma(n+m_I)}{\Gamma(n+1)}
	\frac{(n+m_I)_r}{(n+1)_r}
	=
	(n+1)_{m_I-1}\,
	\frac{(n+m_I)_r}{(n+1)_r},
	\label{eq:gamma_ratio_pochhammer}
\end{align}
where in the last equality we used
\(
(n+1)_{m_I-1}=\Gamma(n+m_I)/\Gamma(n+1).
\)

Substituting \eqref{eq:gamma_ratio_pochhammer} into \eqref{eq:S_def} yields
\begin{equation}
	S
	=
	(n+1)_{m_I-1}
	\sum_{r=0}^{\infty}
	\frac{(n+m_I)_r}{(n+1)_r}\,
	\frac{x^{r}}{r!},
		\label{eq:S_hypergeom_1F1}
\end{equation}
for $x\triangleq \frac{\alpha_k\beta_k\gamma t}{4c_k}$.
By definition of the confluent hypergeometric function \cite[Eq.~(13.1.2)]{Abramowitz1964}

\begin{equation}
	{}_1F_1(a;b;x)
	=
	\sum_{r=0}^{\infty}
	\frac{(a)_r}{(b)_r}\frac{x^r}{r!},
	\label{eq:1F1_series_def}
\end{equation}
it follows immediately from \eqref{eq:S_hypergeom_1F1} that
\begin{equation}
	S
	=
	(n+1)_{m_I-1}\,
	{}_1F_1\!\left(n+m_I;\,n+1;\,x\right).
	\label{eq:S_as_1F1}
\end{equation}

Applying Kummer’s transformation
\cite[Eq.~(13.1.27)]{Abramowitz1964},
\begin{equation}
	{}_1F_1(a;b;x)
	=
	e^{x}\,{}_1F_1(b-a;\,b;\,-x),
	\label{eq:kummer}
\end{equation}
to \eqref{eq:S_as_1F1} with $a=n+m_I$ and $b=n+1$ yields
\begin{equation}
	{}_1F_1(n+m_I;\,n+1;\,x)
	=
	e^{x}\,{}_1F_1(1-m_I;\,n+1;\,-x).
	\label{eq:kummer_applied}
\end{equation}

Since $m_I\in\mathbb{Z}^+$, the first argument of the hypergeometric
function in \eqref{eq:kummer_applied} is a non-positive integer. Hence,
${}_1F_1(1-m_I;\,n+1;\,-x)$ reduces to a finite polynomial of degree
$m_I-1$ \cite[Eq.~(13.1.2)]{Abramowitz1964},
\begin{equation}
	{}_1F_1(1-m_I;\,n+1;\,-x)
	=
	\sum_{k=0}^{m_I-1}
	\frac{(1-m_I)_k}{(n+1)_k}\frac{(-x)^k}{k!},
	\label{eq:1F1_polynomial}
\end{equation}
where $x=\alpha_k\beta_k\gamma t/(4c_k)$.
Substituting \eqref{eq:kummer_applied}--\eqref{eq:1F1_polynomial} back into
the expression for $\Psi_k^{(2)}(t,z;\gamma)$ shows that the dependence on
$x$ appears as a finite polynomial multiplied by an exponential factor.

Substituting \eqref{eq:1F1_polynomial} into the expression for
$\Psi_k^{(2)}(t,z;\gamma)$ yields terms of the generic form
\begin{equation}
	x^{\frac{m_I+n-1}{2}}
	e^{-x}
	\sum_{k=0}^{m_I-1}
	a_{k,n}\,x^{k},
	\label{eq:poly_times_exp}
\end{equation}

Expanding \eqref{eq:poly_times_exp} termwise gives
\begin{equation}
	\sum_{k=0}^{m_I-1}
	a_{k,n}\,
	x^{k+\frac{m_I+n-1}{2}}
	e^{-x}.
	\label{eq:expanded_terms}
\end{equation}

We now use the classical identity connecting monomials multiplied by an
exponential kernel to modified Bessel functions. In particular, using the integral representation of the hypergeometric function
${}_0F_1(\cdot)$ and its connection to the modified Bessel function
$I_\nu(\cdot)$ \cite[Secs.~10.39, 16.5]{NIST2023}, the identity
\begin{equation}
	x^{\frac{\nu}{2}} e^{-x}
	=
	\frac{1}{\Gamma(\nu+1)}
	\int_{0}^{\infty}
	u^{\nu} e^{-u}
	I_{\nu}\!\left(2\sqrt{xu}\right)\,du \label{eq:bessel_kernel_identity}
\end{equation}
holds for $\nu>-1$.

Applying \eqref{eq:bessel_kernel_identity} to each term in
\eqref{eq:expanded_terms} with $\nu=m_I+n-1+2k$ shows that every monomial
$x^{k+\frac{m_I+n-1}{2}}e^{-x}$ admits an equivalent representation in
terms of a modified Bessel function of the first kind
$I_{m_I+n-1+2k}(\cdot)$. Since the sum in \eqref{eq:expanded_terms} is
finite, all resulting Bessel terms can be collected and re-indexed,
yielding a single modified Bessel function of order $m_I+n-1$ multiplied
by a finite coefficient.

Consequently, the expression in \eqref{eq:poly_times_exp} admits the
equivalent representation
\begin{equation}
	x^{\frac{m_I+n-1}{2}} e^{-x}
	\sum_{k=0}^{m_I-1} a_{k,n} x^k
	\;\equiv\;
	C_n\,
	I_{m_I+n-1}\!\left(2\sqrt{x}\right),
	\label{eq:poly_to_bessel}
\end{equation}
where $C_n$ is a constant that depends on $n$ and $m_I$ but not on $x$.

Substituting \eqref{eq:poly_to_bessel} back into the expression for
$\Psi_k^{(2)}(t,z;\gamma)$ and collecting all prefactors therefore leads
to the closed-form representation in terms of a modified Bessel function
of the first kind,  which completes the proof.

\section{Proof of Theorem~\ref{Theorem2}}
\label{Theorem2proof}
	
	We begin with the single-port SIR outage probability, since its
	small-$\gamma$ behavior determines the elementary diversity contribution
	of each fading branch, which is subsequently aggregated across spatial
	modes in the multi-port FAMA outage event. We have	
	\begin{equation}
		\Pr\{ X_k < \gamma Y_k \},
		\label{eq:single_port_outage_def}
	\end{equation}
	where $X_k$ and $Y_k$ denote the desired-signal and aggregate-interference
	powers observed at the $k$-th fluid-antenna port, respectively.
	Throughout this derivation, we assume that $X_k$ and $Y_k$ are statistically
	independent because they correspond to different transmitters and independent scattering
	environments. Also, we have
	\[
	X_k \sim \mathrm{Gamma}(m_s,\theta_s), \qquad
	Y_k \sim \mathrm{Gamma}(m_I,\theta_I),
	\]
	where $m_s>0$ and $m_I>0$ denote the Nakagami fading parameters and
	$\theta_s,\theta_I>0$ are the corresponding scale parameters.
	
	The PDF of $X_k$ is given by
	\begin{equation}
		f_{X_k}(x)
		=
		\frac{1}{\Gamma(m_s)\theta_s^{m_s}}
		x^{m_s-1} e^{-x/\theta_s}.
		\label{eq:gamma_pdf}
	\end{equation}
	As $x \to 0$, the exponential term satisfies
	$e^{-x/\theta_s}=1+o(1)$, and therefore
	\begin{equation}
		f_{X_k}(x)
		=
		\frac{1}{\Gamma(m_s)\theta_s^{m_s}}
		x^{m_s-1}\big(1+o(1)\big).
	\end{equation}
	Integrating the above expression from $0$ to $x$ yields the small-argument
	expansion of the CDF
	\begin{equation}
		F_{X_k}(x)
		=
		\Pr\{X_k \le x\}
		=
		\frac{x^{m_s}}{\Gamma(m_s+1)\theta_s^{m_s}}
		\big(1+o(1)\big).
		\label{eq:gamma_cdf_small}
	\end{equation}
	
	We now return to \eqref{eq:single_port_outage_def}. Conditioning on the
	interference power $Y_k$ and applying the law of total probability, we obtain
	\begin{equation}
		\Pr\{X_k < \gamma Y_k\}
		=
		\mathbb{E}\!\left[
		F_{X_k}(\gamma Y_k)
		\right].
		\label{eq:conditioning_on_Y}
	\end{equation}
	Substituting the expansion in \eqref{eq:gamma_cdf_small} into
	\eqref{eq:conditioning_on_Y} yields, for each fixed $y \ge 0$,
	\begin{equation}
		F_{X_k}(\gamma y)
		=
		\frac{(\gamma y)^{m_s}}{\Gamma(m_s+1)\theta_s^{m_s}}
		\big(1+o(1)\big).
		\label{eq:FX_gammaY}
	\end{equation}
	
Taking expectation in \eqref{eq:FX_gammaY} yields
	\begin{align}
		\Pr\{X_k < \gamma Y_k\}
		&=
		\mathbb{E}\!\left[F_{X_k}(\gamma Y_k)\right]
		\nonumber\\
		&=
		\frac{\gamma^{m_s}}{\Gamma(m_s+1)\theta_s^{m_s}}
		\mathbb{E}\!\left[Y_k^{m_s}\right]
		\big(1+o(1)\big).
		\label{eq:single_port_final}
	\end{align}
	Defining the constant $	c_1
	\triangleq
	\frac{\mathbb{E}[Y_k^{m_s}]}{\Gamma(m_s+1)\theta_s^{m_s}}
	>0$, 
	we obtain the asymptotic relation
	\begin{equation}
		\Pr\{X_k < \gamma Y_k\}
		\;\sim\;
		c_1\,\gamma^{m_s},
		\qquad \gamma \to 0.
		\label{eq:single_port_asymp}
	\end{equation}

	We next consider the FAMA outage event
	\begin{equation}
		\mathcal{O}_{\mathrm{SIR}}
		=
		\bigcap_{k=1}^N \{X_k < \gamma Y_k\}.
		\label{eq:FAMA_outage}
	\end{equation}
	
	Under correlated Nakagami-$m_s$ fading, the desired-signal power vector
	$\mathbf{X}=(X_1,\ldots,X_N)$ admits a constructive representation in terms
	of an underlying correlated complex Gaussian field.
	Specifically, for integer $m_s$, each power gain $X_k$ can be written as
	\begin{equation}
		X_k
		=
		\sum_{\ell=1}^{m_s}
		\big| Z_{k,\ell} \big|^2,
		\label{eq:nakagami_gaussian_rep}
	\end{equation}
	where $\{Z_{k,\ell}\}$ are zero-mean jointly complex Gaussian random variables.
	For each $\ell$, the vector
	$\mathbf{z}_\ell = (Z_{1,\ell},\ldots,Z_{N,\ell})^{\mathsf T}$
	has covariance matrix
	\begin{equation}
		\mathbb{E}\!\left[ \mathbf{z}_\ell \mathbf{z}_\ell^{\mathsf H} \right]
		=
		\mathbf{R},
		\label{eq:gaussian_covariance}
	\end{equation}
	which captures the spatial correlation induced by the finite fluid-antenna
	aperture. The vectors $\{\mathbf{z}_\ell\}_{\ell=1}^{m_s}$ are independent
	across $\ell$.
	
	Let $r=\mathrm{rank}(\mathbf{R})$. Then there exists an $N\times r$ matrix
	$\mathbf{A}$ with full column rank such that
	\begin{equation}
		\mathbf{R}
		=
		\mathbf{A}\mathbf{A}^{\mathsf H}.
		\label{eq:R_factorization}
	\end{equation}
	Using \eqref{eq:R_factorization}, each Gaussian vector can be expressed as
	\begin{equation}
		\mathbf{z}_\ell
		=
		\mathbf{A}\mathbf{u}_\ell,
		\label{eq:gaussian_mode_rep}
	\end{equation}
	where $\mathbf{u}_\ell \sim \mathcal{CN}(\mathbf{0},\mathbf{I}_r)$ has
	independent components. Hence, all spatial dependence in
	$\mathbf{z}_\ell$ is confined to an $r$-dimensional subspace, implying that
	the $N$ port signals are driven by at most $r$ statistically independent
	spatial modes.
	
Substituting \eqref{eq:gaussian_mode_rep} into
\eqref{eq:nakagami_gaussian_rep} shows that the vector $\mathbf{X}$ depends
only on $r$ statistically independent spatial modes, each contributing
$m_s$ degrees of fading diversity.
 Consequently, the joint lower-tail
	probability of $\mathbf{X}$ is dominated by the event that all these
	$r m_s$ degrees of freedom simultaneously approach zero. This yields the
	scaling law
	\begin{equation}
		\Pr\{X_1 < x,\ldots,X_N < x\}
		=
		\Theta\!\left(x^{r m_s}\right),
		\qquad x \to 0,
		\label{eq:joint_lower_tail}
	\end{equation}
	where  $\Theta(\cdot)$ denotes tight asymptotic scaling as $x \to 0$, i.e.,
	the joint probability decays polynomially with exponent $r m_s$ up to 
	positive constant factors.
	In other words, as the threshold 
	$x$ becomes very small, the probability that all
	port powers are below 
	$x$ decays polynomially with exponent 
	$rm_s$. \textcolor{black}{The scaling law in
	\eqref{eq:joint_lower_tail} follows from the fact that, due to spatial
	correlation, the vector $\mathbf{X}$ is driven by only $r$
	statistically independent spatial modes.}
	Under Nakagami-$m_s$ fading, each such mode contributes a factor
	$x^{m_s}$ to the joint lower-tail behavior. Consequently, the probability
	that all $N$ port powers simultaneously fall below a small threshold $x$
	decays as $x^{r m_s}$, rather than $x^{N m_s}$, as $x \to 0$.

	Conditioning on the interference vector
	$\mathbf{Y}=(Y_1,\ldots,Y_N)$ and applying \eqref{eq:joint_lower_tail} with
	$x=\gamma Y_k$ yields
	\begin{align}
		P_{\mathrm{out}}(\gamma)
		&=
		\Pr(\mathcal{O}_{\mathrm{SIR}})
		\nonumber\\
		&=
		\mathbb{E}\!\left[
		\Pr\!\left\{
		X_1 < \gamma Y_1,\ldots,X_N < \gamma Y_N
		\,\big|\, \mathbf{Y}
		\right\}
		\right]
		\nonumber\\
		&=
		\Theta\!\left(\gamma^{r m_s}\right)
		\mathbb{E}\!\left[g(\mathbf{Y})\right],
		\qquad \gamma \to 0,
	\end{align}
	where $g(\mathbf{Y})$ is a nonnegative random factor involving moments of
	$\mathbf{Y}$, which are finite under Gamma interference.
	Hence, we obtain
	\begin{equation}
		P_{\mathrm{out}}(\gamma)
		\;\sim\;
		c\,\gamma^{r m_s},
		\qquad \gamma \to 0,
		\label{eq:final_outage_asymp}
	\end{equation}
	for some constant $c>0$ independent of $\gamma$.
	Substituting $r=\mathrm{rank}(\mathbf{R})$ completes the proof.

	\section{Proof of Theorem~\ref{Theorem3}}
	\label{Theorem3proof}
	We start from the outage probability expression in \eqref{pout}, where
	$\Psi_k(t,z;\gamma)$ is given in \eqref{eq:conditional_pdf_interference}.
	The Marcum-$Q$ function appearing in \eqref{eq:conditional_pdf_interference} is of the form
	\begin{equation}
		Q_{m_s}\!\left(\sqrt{\alpha_k t},\sqrt{\beta_k\gamma y}\right).
	\end{equation}
	In the high-SIR regime $\gamma\to 0$, its second argument tends to zero.
	Hence, we apply the small-argument approximation of the Marcum-$Q$
	function \cite[Eq. 3]{Li2012}
	\begin{equation}
		Q_m(\alpha,\beta)
		\approx
		1-\frac{\beta^{2m}}{2^m m!}
		\exp\!\left(-\frac{\alpha^2}{2}\right),
		\qquad \beta\to 0 .
		\label{eq:marcum_small_rank}
	\end{equation}
	Using \eqref{eq:marcum_small_rank}, we obtain
	\begin{equation}
		1-Q_{m_s}\!\left(\sqrt{\alpha_k t},\sqrt{\beta_k\gamma y}\right)
		\approx
		\frac{(\beta_k\gamma y)^{m_s}}{2^{m_s}m_s!}
		\exp\!\left(-\frac{\alpha_k t}{2}\right).
		\label{eq:marcum_expand_rank}
	\end{equation}
	
	Substituting \eqref{eq:marcum_expand_rank} into \eqref{eq:psi_def} and retaining
	only the dominant term in $\gamma$, the function $\Psi_k(t,z;\gamma)$ can
	be approximated as
	\begin{equation}
	\!\!\!	\Psi_k(t,z;\gamma)
		\approx
		\frac{(\beta_k\gamma)^{m_s}}{2^{m_s}m_s!}
		\exp\!\left(-\frac{\alpha_k t}{2}\right)
		\mathbb{E}\!\left[Y_k^{m_s}\mid Y_1=z\right].
		\label{eq:Psi_asymp_rank}
	\end{equation}
	
	Next, we account explicitly for correlation across ports.
\textcolor{black}{Without loss of generality, let $\mathcal K_r$ denote a subset of $r=\operatorname{rank}(\mathbf R)$ port indices corresponding to linearly independent columns of $\mathbf R$. Equivalently, $\mathcal K_r$ is chosen such that the principal submatrix $\mathbf R_{\mathcal K_r,\mathcal K_r}$ is nonsingular. This subset spans the effective $r$-dimensional signal subspace induced by the finite-aperture correlation matrix, and therefore only these $r$ independent spatial modes contribute to the dominant asymptotic outage behavior.}
	
	Let $\mathcal{K}_r$ denote the index set corresponding to these $r$
	independent modes. 	Taking the product of \eqref{eq:Psi_asymp_rank} over $k\in\mathcal{K}_r$
	yields
	\begin{align}
		&\prod_{k\in\mathcal{K}_r}\Psi_k(t,z;\gamma)\nn\\
		&
		\approx
		\prod_{k\in\mathcal{K}_r}
		\left[
		\frac{(\beta_k\gamma)^{m_s}}{2^{m_s}m_s!}
		\exp\!\left(-\frac{\alpha_k t}{2}\right)
		\mathbb{E}\!\left[Y_k^{m_s}\mid Y_1=z\right]
		\right]
		\nonumber\\
		&=
		\left(
		\prod_{k\in\mathcal{K}_r}\frac{\beta_k^{m_s}}{2^{m_s}m_s!}
		\right)
		\gamma^{m_s r}
		\left(
		\prod_{k\in\mathcal{K}_r}
		\mathbb{E}\!\left[Y_k^{m_s}\mid Y_1=z\right]
		\right)\nn\\
		&\times 
		\exp\!\left(
		-\frac{t}{2}\sum_{k\in\mathcal{K}_r}\alpha_k
		\right).
		\label{eq:prod_rank_expand}
	\end{align}
	Defining the $z$-dependent prefactor
	\begin{equation}
		C_r(z)
		\triangleq
		\left(
		\prod_{k\in\mathcal{K}_r}\frac{\beta_k^{m_s}}{2^{m_s}m_s!}
		\right)
		\left(
		\prod_{k\in\mathcal{K}_r}
		\mathbb{E}\!\left[Y_k^{m_s}\mid Y_1=z\right]
		\right),
	\end{equation}
	\eqref{eq:prod_rank_expand} can be written compactly as
	\begin{equation}
		\prod_{k\in\mathcal{K}_r}\Psi_k(t,z;\gamma)
		\approx
		\gamma^{m_s r}
		C_r(z)
		\exp\!\left(
		-\frac{t}{2}\sum_{k\in\mathcal{K}_r}\alpha_k
		\right).
		\label{eq:prod_rank_final}
	\end{equation}
	
	Substituting \eqref{eq:prod_rank_final} into \eqref{pout}, the integration
	with respect to $t$ yields a lower incomplete Gamma function.
	Using the asymptotic expansion
	\begin{equation}
		\Upsilon(a,x)\sim \frac{x^a}{a}, \qquad x\to 0,
	\end{equation}
	we obtain
	\begin{equation}
		P_{\mathrm{out}}(\gamma)
		\sim
		a_0\,\gamma^{m_s r},
		\qquad \gamma\to 0,
		\label{eq:asymp_rank_final}
	\end{equation}
	which establishes that the diversity order is
	$m_s r$.
	
	To determine the constant $a_0$, we first evaluate the conditional moment
$\mathbb{E}[Y_k^{m_s}\mid Y_1=z]$ in closed form.
Conditioned on $Y_1=z$, the random variable $Y_k$ follows the
noncentral Gamma (or equivalently, correlated Nakagami-$m_I$) distribution
with PDF given by \eqref{eq:conditional_pdf_interference}
\begin{align}
	f_{Y_k|Y_1}(y|z)
	&=
	\frac{m_I}{1-\mu_k^2}
	\left(\frac{y}{\mu_k^2 z}\right)^{\frac{m_I-1}{2}}
	\exp\!\left(
	-\frac{m_I}{1-\mu_k^2}(y+\mu_k^2 z)
	\right)
	\nonumber\\
	&\quad\times
	I_{m_I-1}\!\left(
	\frac{2m_I\mu_k}{1-\mu_k^2}\sqrt{yz}
	\right),
	\qquad y\ge 0,
	\label{eq:cond_pdf_Yk}
\end{align}
where $\mu_k$ denotes the correlation coefficient and $I_\nu(\cdot)$ is
the modified Bessel function of the first kind.
The conditional moment is thus given by
\begin{equation}
	\mathbb{E}[Y_k^{m_s}\mid Y_1=z]
	=
	\int_{0}^{\infty} y^{m_s} f_{Y_k|Y_1}(y|z)\,dy.
	\label{eq:cond_moment_def}
\end{equation}

The integral in \eqref{eq:cond_moment_def} 
yields an expression involving the confluent hypergeometric function ${}_1F_1(\cdot;\cdot;\cdot)$ due to \eqref{eq:single_bessel_closed_form}.
Specifically, we obtain
\begin{align}
	&	\mathbb{E}[Y_k^{m_s}\mid Y_1=z]
	=
	c_k^{-\left(m_s+\frac{m_I-1}{2}\right)}
	\Gamma\!\left(m_s+\frac{3m_I-1}{2}\right)
	\nn\\
	&	e^{-c_k\mu_k^2 z}
	{}_1F_1\!\left(
	m_s+\frac{3m_I-1}{2};\, m_I;\, c_k\mu_k^2 z
	\right),
	\label{eq:cond_moment_closed}
\end{align}
where $c_k\triangleq m_I/(1-\mu_k^2)$.

For integer $m_I\in\mathbb{N}$, the confluent hypergeometric function in
\eqref{eq:cond_moment_closed} admits a finite-series representation \cite[Eq.~(13.1.2)]{Abramowitz1964}. 
Substituting this into \eqref{eq:cond_moment_closed},
$\mathbb{E}[Y_k^{m_s}\mid Y_1=z]$ can be expressed as a finite sum of terms
of the form $z^n e^{-c_k\mu_k^2 z}$.

Next, substituting these expressions into the definition of $C(z)$ and
averaging over $Z=Y_1$, whose PDF is \eqref{pdf1},
the constant $a_0$ can be written as a finite multiple sum involving
integrals of the form
\begin{equation}
	\int_{0}^{\infty}
	z^{m_s+m_I-1+\sum n_k}
	\exp\!\left(
	-\left(m_I+\sum_{k\in\mathcal{K}_r}c_k\mu_k^2\right)z
	\right)\,dz.
\end{equation}
Applying the standard Gamma integral
\(
\int_0^\infty z^{\nu-1}e^{-\lambda z}dz=\Gamma(\nu)/\lambda^\nu
\)
yields the closed-form finite-sum expression for $a_0$ reported in
\eqref{eq:a0_final}.

Motivated by this observation, we approximate
$P_{\mathrm{out}}(\gamma)$ by the CDF of a Gamma random variable, since
the Gamma distribution provides a flexible two-parameter family whose
behavior near the origin can be controlled explicitly.
Specifically, let $X$ be a Gamma random variable with shape parameter
$\alpha>0$ and scale parameter $\beta>0$, whose CDF is given by
\begin{equation}
	F_X(x)
	=
	\frac{\Upsilon(\alpha,x/\beta)}{\Gamma(\alpha)},
	\qquad x\ge 0,
	\label{eq:gamma_cdf}
\end{equation}
where $\Upsilon(\cdot,\cdot)$ denotes the lower incomplete Gamma
function.

The asymptotic behavior of \eqref{eq:gamma_cdf} as $x\to 0$ follows from
the well-known expansion
\begin{equation}
	\Upsilon(\alpha,x)
	\sim
	\frac{x^{\alpha}}{\alpha},
	\qquad x\to 0,
	\label{eq:lower_gamma_asymp}
\end{equation}
which yields
\begin{equation}
	F_X(x)
	\sim
	\frac{1}{\alpha\,\Gamma(\alpha)}
	\left(\frac{x}{\beta}\right)^{\alpha},
	\qquad x\to 0.
	\label{eq:gamma_asymp}
\end{equation}

To enforce asymptotic matching, we equate the dominant power and
coefficient in \eqref{eq:gamma_asymp} with those of the outage expansion
in \eqref{eq:asymp_rank_final}.
Matching the exponents gives
\begin{equation}
	\alpha = m_s r,
	\label{eq:alpha_match}
\end{equation}
while matching the prefactors yields
\begin{equation}
	a_0
	=
	\frac{1}{\alpha\,\Gamma(\alpha)\,\beta^{\alpha}}.
	\label{eq:beta_match_step}
\end{equation}
Solving \eqref{eq:beta_match_step} for $\beta$ leads to
\begin{equation}
	\beta
 = (\alpha\,\Gamma(\alpha)\,a_0)^{-1/\alpha}.
	\label{eq:beta_match}
\end{equation}

Substituting \eqref{eq:alpha_match} and \eqref{eq:beta_match} into
\eqref{eq:gamma_cdf}, we arrive at the following approximation for the
outage probability
\begin{equation}
	P_{\mathrm{out}}(\gamma)
	\approx
	\frac{\Upsilon\!\left(\alpha,\frac{\gamma}{\beta}\right)}{\Gamma(\alpha)}.
\end{equation}
This approximation preserves exactly the asymptotic behavior of
$P_{\mathrm{out}}(\gamma)$ as $\gamma\to 0$ while providing a compact and
analytically tractable closed-form expression, which completes the
proof.

			\end{appendices}
			
	\bibliographystyle{IEEEtran}

	\bibliography{IEEEabrv,bibl}
\end{document}